\documentclass[acmsmall]{acmart}
\AtBeginDocument{%
  }

\copyrightyear{2026}

\setcopyright{cc}
\setcctype{by}
\acmJournal{POMACS}
\acmYear{2026} \acmVolume{10} \acmNumber{2} \acmArticle{55}
\acmMonth{6} \acmDOI{10.1145/3805653}

\usepackage{algorithm}
\usepackage{algpseudocode}
\usepackage{graphicx}
\usepackage{subcaption}
\usepackage{textcomp}
\usepackage{xcolor}
\usepackage{tabularx}
\usepackage{xspace}
\usepackage{pifont}
\usepackage{multirow}
\usepackage{makecell}
\usepackage{booktabs}
\usepackage{array}
\usepackage{amsmath,amsthm}
\usepackage{bbm}  
\newcolumntype{C}[1]{>{\centering\arraybackslash}m{#1}}
\newcommand{\acronym}{\textit{WISP}\xspace}
\newcommand{\acronymsled}{\textit{SLED}\xspace}
\renewcommand{\shortauthors}{Li et al.}

\newtheorem{theorem}{Theorem}        

\begin{document}

\title{\acronym: Waste- and Interference-Suppressed Distributed Speculative LLM Serving at the Edge via Dynamic Drafting and SLO-Aware Batching}

\author{Xiangchen Li}
\orcid{0000-0002-5293-5348}
\affiliation{%
  \institution{Virginia Tech}
  \city{Blacksburg}
  \state{Virginia}
  \country{USA}
}
\email{lixiangchen@vt.edu}

\author{Jiakun Fan}
\affiliation{%
  \institution{Virginia Tech}
  \city{Blacksburg}
  \state{Virginia}
  \country{USA}}
\email{jiakunfan@vt.edu}

\author{Qingyuan Wang}
\affiliation{%
  \institution{University College Dublin}
  \city{Dublin}
  \country{Ireland}}
\email{qingyuan.wang@ucdconnect.ie}

\author{Dimitrios Spatharakis}
\affiliation{%
  \institution{National Technical University of Athens}
  \city{Athens}
  \country{Greece}
}
\email{dspatharakis@netmode.ntua.gr}

\author{Saeid Ghafouri}
\affiliation{%
 \institution{Queen's University Belfast}
 \city{Belfast}
 \state{Northern Ireland}
 \country{UK}}
\email{s.ghafouri@qub.ac.uk}

\author{Hans Vandierendonck}
\affiliation{%
  \institution{Queen's University Belfast}
  \city{Belfast}
  \state{Northern Ireland}
  \country{UK}}
\email{h.vandierendonck@qub.ac.uk}

\author{Deepu John}
\affiliation{%
  \institution{University College Dublin}
  \city{Dublin}
  \country{Ireland}}
\email{deepu.john@ucd.ie}

\author{Bo Ji}
\affiliation{%
  \institution{Virginia Tech}
  \city{Blacksburg}
  \state{Virginia}
  \country{USA}}
\email{boji@vt.edu}

\author{Ali R. Butt}
\affiliation{%
  \institution{Virginia Tech}
  \city{Blacksburg}
  \state{Virginia}
  \country{USA}}
\email{butta@cs.vt.edu}

\author{Dimitrios~S. Nikolopoulos}
\affiliation{%
  \institution{Virginia Tech}
  \city{Blacksburg}
  \state{Virginia}
  \country{USA}}
\email{dsn@vt.edu}

\renewcommand{\shortauthors}{Li et al.}

\begin{abstract}
  As Large Language Models (LLMs) become increasingly accessible to end users, an ever-growing number of inference requests are initiated from edge devices and computed on centralized GPU clusters. However, the resulting exponential growth in computation workload is placing significant strain on data centers, while edge devices remain largely underutilized, leading to imbalanced workloads and resource inefficiency across the network. Integrating edge devices into the LLM inference process via speculative decoding helps balance the workload between the edge and the cloud, while maintaining lossless prediction accuracy. In this paper, we identify and formalize two critical bottlenecks that limit the efficiency and scalability of distributed speculative LLM serving: \emph{Wasted Drafting Time} and \emph{Verification Interference}. To address these challenges, we propose \acronym, an efficient and SLO-aware distributed LLM inference system that consists of an intelligent speculation controller, a verification time estimator, and a verification batch scheduler. These components collaboratively enhance drafting efficiency and optimize verification request scheduling on the server. Extensive numerical results show that \acronym improves system capacity by up to 2.1× and 4.1×, and increases system goodput by up to 1.94× and 3.7×, compared to centralized serving and \acronymsled, respectively.
\end{abstract}

\ccsdesc[500]{Computing methodologies~Distributed computing methodologies}
\ccsdesc[500]{Computing methodologies~Distributed artificial intelligence}
\ccsdesc[300]{Computer systems organization~Distributed architectures}
\ccsdesc[300]{Computing methodologies~Natural language processing}
\ccsdesc[500]{Networks~Cloud computing}
\ccsdesc[500]{General and reference~General conference proceedings}

\keywords{Speculative Decoding, Large Language Models, Edge Computing, Distributed Inference, Token Verification, Resource-Aware Serving}

\received{14 January 2026}
\received[revised]{27 March 2026}
\received[accepted]{6 April 2026}

\maketitle

\section{Introduction}
Large Language Models (LLMs) have revolutionized the landscape of artificial intelligence, demonstrating unprecedented capabilities in natural language understanding, reasoning, and content generation. However, the widespread adoption of LLMs is severely constrained by the prohibitive costs associated with inference, making LLM inference a pure GPU cluster workload while leaving user devices idle. The exponentially growing demand for LLM inference, particularly on edge devices such as personal assistants and in-field systems, has intensified the GPU Floating Point Operations Per Second (FLOPS) crisis, making the computational power of these devices increasingly indispensable.

Unlike traditional machine learning workloads, which are typically system-initiated, batch-style prediction services processed within centralized data centers, LLMs are directly exposed to end users as interactive assistants and are triggered at the network edge by smartphones, personal computers, and IoT devices. ChatGPT processes roughly 2.5 billion user prompts per day~\cite{axios2025chatgpt}, Apple reports that Siri handles approximately 1.5 billion voice requests daily~\cite{apple2024siri}, and Google’s Gemini ecosystem serves roughly 650 million monthly users via the Gemini app and nearly 2 billion monthly users through AI Overviews in Search~\cite{alphabet2025q3, alphabet2025q2}. Despite this massive volume of inference initiated at the edge, the computation is still predominantly centralized in data centers, with the majority of LLM inference handled by GPU clusters. This centralization creates a significant resource inefficiency: while billions of edge devices—now increasingly equipped with capable NPUs and GPUs—sit largely idle during inference requests, the cloud bears the full computational burden, incurring high operational costs.

In contrast, on-device LLM inference~\cite{zhang2024edgeshard, yu2024edge} seeks to exploit device-side computational resources by deploying lightweight or compressed models capable of token generation under resource-constrained environments. However, the resulting output quality remains inadequate in today’s competitive landscape, where model capacity plays a critical role in performance.

To reconcile the need for utilizing edge FLOPS with the demand for competitive output quality, distributed speculative edge inference systems offer a promising direction to achieve both efficiency and performance. The "drafting-then-verifying" paradigm introduced by speculative decoding \cite{leviathan2023fast} enables a collaborative edge-cloud approach, where token generation is offloaded to underutilized edge resources while verification is retained in the cloud. Consequently, more GPU resources are conserved for efficient verification rather than being consumed by token generation. Specifically, SLED \cite{li2025sled} operationalized this idea by orchestrating speculative execution across heterogeneous edge devices and servers. Nevertheless, SLED remains a preliminary framework with critical limitations; most notably, its reliance on static batching and best-effort scheduling fails to guarantee strict Service Level Objectives (SLOs), resulting in unpredictable latency in real-world, multi-tenant scenarios.

In this paper, we present \acronym, a SLO-aware, efficient, and intelligent LLM inference system based on speculative decoding for heterogeneous devices. Building on the distributed speculative edge LLM serving architecture, we identify and formally define \emph{Wasted Drafting Time} (WDT), the cumulative latency spent on edge-side draft tokens that are later rejected by the server, which directly degrades device throughput and overall system capacity. To reduce WDT, \acronym introduces an intelligent drafting controller that triggers verification dynamically using a lightweight reject predictor, which is mathematically proved to reduce WDT. We further incorporate SLOs in terms of token generation speed to capture heterogeneous application requirements and identify \emph{Verification Interference}, which is the head-of-line effect induced by mixing heterogeneous verification requests within a GPU micro-batch, as a primary cause of SLO violations. To mitigate interference, we formulate SLO-aware batch construction as a knapsack problem and design a greedy scheduler that accounts for both SLO slack and interference. Finally, \acronym employs a fine-grained verification-time estimator, profiled across batch configurations, to enable online feasibility checks during scheduling. Extensive simulations and experiments show that \acronym enlarges the feasible operating region of speculative edge serving: under token-speed SLOs, it improves system capacity by up to $2.1\times$ over centralized serving and $4.1\times$ over \acronymsled, and increases end-to-end system goodput by up to $1.94\times$ and $3.7\times$, respectively, while markedly reducing SLO violations under multi-tenant verification interference.

Our key contributions are summarized as follows:
\begin{itemize}
    \item For the first time, we identify and define two fundamental bottlenecks in distributed speculative edge LLM serving: \emph{Wasted Drafting Time} (WDT) and \emph{Verification Interference}, and we provide extensive evaluation showing that they fundamentally limit scalability by driving SLO violations and suppressing throughput/capacity. Guided by these insights, we propose \acronym, an SLO-aware collaborative LLM inference system that distributes inference across user devices and GPU clusters to reduce violations while improving device throughput and overall system capacity.
    
    \item We develop a predictive modeling framework to orchestrate efficient collaborative execution, including an ML-based rejection predictor (mathematically proved to reduce WDT) and a fine-grained verification-time estimator that predicts server-side batch latency by separately modeling self-attention and linear-operation costs.
    
    \item We formulate SLO-aware batch construction as a knapsack problem and propose a greedy scheduling algorithm that explicitly accounts for SLO slack and \emph{Verification Interference}.
    
    \item Finally, we build an end-to-end evaluation pipeline, including trace-driven simulations and a prototype system, and show that \acronym consistently improves system goodput/throughput and capacity while significantly reducing SLO violations under verification interference, compared with centralized serving and prior speculative edge baselines.
\end{itemize}

The remainder of this paper is organized as follows: Section \ref{sec:related_work} categorizes and discusses existing thrusts of pushing LLM towards edge devices into on-device inference, model splitting-based inference, and distributed speculative LLM serving system, and the basic theory of speculative decoding is also covered. Section \ref{sec:motivations} identifies and evaluates the two key factors in the existing collaborative speculative decoding-based inference systems, which also motivate the system design of this work. Then, the \acronym is introduced and elaborated in Section \ref{sec:sled}. Later, Section \ref{sec:experiments} shows the settings and results of numerical simulations and experiments. Finally, Section \ref{sec:conclusion} concludes this paper.

\section{Related Work and Background}\label{sec:related_work}
The rapidly evolving landscape of edge-based Large Language Model (LLM) serving can be broadly categorized into three distinct technical thrusts: pure on-device inference, model splitting inference (partitioning layers between device and cloud), and collaborative inference frameworks that leverage speculative decoding to orchestrate resources across the edge-cloud continuum.

\subsection{On-device Inference}
Pure on-device inference aims to execute the entire model on the edge device to ensure maximum privacy and offline availability. To achieve this, researchers have primarily focused on model compression techniques, such as quantization and pruning, or the development of specialized lightweight architectures like MobileLLM~\cite{liu2024mobilellm}. Memory-efficient inference has also been explored; notably, LLM in a Flash~\cite{alizadeh2024llm} enables running models up to twice the size of available DRAM by strategically loading parameters from flash memory on demand. For heterogeneous hardware utilization, PowerInfer~\cite{song2024powerinfer} exploits activation sparsity to design a GPU-CPU hybrid inference engine, achieving up to 11.69$\times$ speedup over llama.cpp on consumer-grade GPUs, while its successor PowerInfer-2~\cite{xue2024powerinfer} extends this approach to smartphones. Addressing the scheduling bottlenecks in such hybrid setups, APEX~\cite{fan2025parallel} introduces a profiling-informed scheduling strategy that dynamically overlaps CPU self-attention operations with GPU to maximize parallelism on memory-constrained hardware. More recently, system-level optimizations have emerged; for instance, EdgeLLM~\cite{yu2024edge} implements a complete speculative decoding framework directly on the device, utilizing a self-adaptive fallback strategy to manage resource consumption. Lightweight inference engines such as llama.cpp~\cite{llama.cpp} and MLC-LLM~\cite{mlc-llm} have also enabled practical deployment of quantized models on edge platforms. However, despite these advancements, the viability of pure on-device inference remains fundamentally bound by hardware physics. At the end of the day, the performance and quality of on-device inference are strictly constrained by the actual size of the model that can be squeezed into the device's limited DRAM and power budget, often precluding the deployment of state-of-the-art multi-billion parameter models.

\subsection{Model Splitting}
Another approach to leverage edge resources is model splitting, where the heavy LLM is partitioned across the edge device and the cloud server to distribute the computational load. A prominent direction in this domain is \textit{phase disaggregation}, which isolates the distinct computational patterns of LLM inference. For instance, Splitwise \cite{patel2024splitwise} and P/D-Device \cite{jin2025p} decouple the prefill and decoding phases, offloading the compute-intensive prompt processing to the cloud while executing the latency-sensitive token generation on the edge. Beyond phase splitting, other works focus on adaptive layer partitioning; EdgeShard \cite{zhang2024edgeshard} employs dynamic programming to jointly optimize device selection and model partitioning across heterogeneous clusters. Similarly, SplitLLM \cite{mudvari2024splitllm} develops efficient splitting algorithms that balance computation and communication costs to maximize throughput. To adapt to fluctuating environments, Galaxy \cite{ye2024galaxy} and other learning-based frameworks \cite{chen2024adaptive} propose resource-efficient coordination protocols to maintain consistency across distributed nodes. However, despite their theoretical benefits, these pipelining strategies necessitate the frequent transmission of large intermediate activation tensors. The resulting synchronization overhead and the bandwidth bottlenecks of wireless networks often negate the computational gains, making them challenging to scale for real-time interactive applications.

\subsection{Collaborative edge LLM serving system}
Since SLED \cite{li2025sled} first introduced speculative decoding into the collaborated LLM inference for edge devices, more and more attention has been paid to this more promising technique. For instance, Venkatesha et al. \cite{venkatesha2025fast} addressed the challenge of heterogeneous vocabularies between draft and target models, proposing a universal speculative decoding mechanism that enables flexible model pairing without retraining. However, their work primarily focuses on the algorithmic compatibility of model pairs and overlooks the system-level scheduling complexities required to support multiple concurrent edge users under resource constraints. More recently, Yu et al. \cite{yu2025dsd} proposed DSD, a distributed speculative decoding framework that introduces an Adaptive Window Control (AWC) policy to dynamically adjust speculation window sizes for throughput optimization. While effective for single-stream performance, DSD optimizes primarily for average throughput rather than strict tail latency, failing to provide the explicit Service Level Objective (SLO) guarantees necessary for real-time interactive applications. Similarly, Ning et al. \cite{ning2025dssd} introduced Distributed Split Speculative Decoding (DSSD), which partitions the verification phase between device and edge to minimize the communication overhead of transmitting vocabulary distributions. Although this reduces uplink payload, the rigid partitioning of the verification logic introduces high synchronization dependencies and does not address the server-side contention when scaling to massive crowds of heterogeneous edge devices.

\subsection{Background: Speculative Decoding}
\label{sec:related_work:background}
Speculative decoding accelerates autoregressive generation by combining a light-weight \emph{draft} model and a high-quality \emph{target} model. Given the current prefix $\mathbf{x}$, the draft model sequentially proposes a block of $K$ draft tokens $\mathbf{y}_{1:K}\triangleq (y_1,\dots,y_K)$, where $y_i \sim q(\cdot \mid \mathbf{x}, \mathbf{y}_{<i})$ and $q(\cdot)$ denotes the draft next-token distribution. The target model then verifies these draft tokens in order using the standard acceptance rule:
\begin{equation}
a_i \triangleq \min\left\{1,\ \frac{p\!\left(y_i \mid \mathbf{x}, \mathbf{y}_{<i}\right)}{q\!\left(y_i \mid \mathbf{x}, \mathbf{y}_{<i}\right)}\right\},
\qquad
A_i \triangleq \mathbbm{1}\!\left[u_i \le a_i\right],\ \ u_i \sim \mathrm{Unif}(0,1),
\label{eq:spec:accept}
\end{equation}
where $p(\cdot)$ is the target next-token distribution and $A_i\in\{0,1\}$ indicates whether the $i$-th draft token is accepted. Let
\begin{equation}
R \triangleq \min\{i \in \{1,\dots,K\}: A_i=0\},\ \ \text{with } R=K+1 \text{ if all are accepted}, 
\qquad
L \triangleq R-1,
\label{eq:spec:R-L}
\end{equation}
so that $\mathbf{y}_{1:L}$ are accepted and the process stops at position $R$. Finally, the target model produces one \emph{target token} at the stopping position,
\begin{equation}
\tilde{y}_R \sim p(\cdot \mid \mathbf{x}, \mathbf{y}_{<R}),
\label{eq:spec:target-token}
\end{equation}
which replaces the first rejected draft token when $R\le K$, or becomes the next token after a fully accepted block when $R=K+1$. In this way, each drafting-verifying iteration appends $\mathbf{y}_{1:L}$ and $\tilde{y}_R$ to the prefix $\mathbf{x}$ while preserving the target distribution. By reducing the forwarding requests on the large target model, speculative decoding is capable of generating tokens faster and more efficient than traditional autoregressive decoding.

\section{Motivations} \label{sec:motivations}
\subsection{SLO-aware Inference}
\label{sec:motivation:slo-aware}

Edge applications exhibit heterogeneous responsiveness expectations. Workloads that involve continuous text consumption, such as conversational interaction or document reading, typically benefit from steady token streaming. Human studies report an average silent reading rate of about 175--300 words per minute, which corresponds to a moderate token streaming rate on the order of $4$--$6$ tokens/s for fluent interaction \cite{brysbaert2019readingrate}. In contrast, programming-oriented tasks may demand higher token rates to match the semantic density of code \cite{bibaev2022all}. Conversely, in decision- or command-driven settings, such as industrial operator assistance, responses are often deliberately constrained to short outputs (e.g., one or two sentences), and can therefore tolerate lower token generation speeds \cite{freire2023factories}.  

Prior distributed speculative edge-serving systems largely treat the verification server as a best-effort, or First Come First Serve (FCFS) backend \cite{li2025sled}, without explicitly reasoning about its SLOs. However, in speculative edge serving, user-perceived responsiveness is ultimately bounded by how quickly the verification server can \emph{return and commit} results. Therefore, supporting consistent responsiveness across heterogeneous edge applications requires the verification server to be SLO-aware, rather than operating in a purely best-effort manner.

Table~\ref{tab:fcfs_hetero_devices} reports the token-generation-speed SLO violation behavior of an FCFS verification system in a heterogeneous-speed setting. 
In this experiment, multiple edge devices issue requests with different target token speeds, while verification is served by a single GPU verifier running the Qwen3-32B target model on an NVIDIA A100-80GB. 
The results indicate that FCFS becomes increasingly incapable of meeting stringent speed targets as more devices share the verifier, whereas only the most relaxed targets remain consistently achievable. 
This is because FCFS is SLO-agnostic: by treating requests with different urgency uniformly, it amplifies queueing and contention, causing time-sensitive requests to be delayed by less urgent traffic and motivating SLO-aware verification scheduling.

Motivated by these practical variations, \acronym groups requests into $C$ SLO classes and assigns each class $c \in {1,2,\dots,C}$ a target token generation speed $s_c$ (tokens/s).

\begin{table}[t]
\centering
\scriptsize
\setlength{\tabcolsep}{6pt}
\renewcommand{\arraystretch}{1.15}
\caption{Token Generation Speed Violation Rate under FCFS Verification Strategy.}
\label{tab:fcfs_hetero_devices}
\begin{tabular}{lcccc}
\toprule
\textbf{System} &
\textbf{Class 4: 8 (tokens/s)} &
\textbf{Class 3: 6 (tokens/s)} &
\textbf{Class 2: 4 (tokens/s)} &
\textbf{Class 1: 2 (tokens/s)} \\
\midrule
\acronym & 98.27\% & 82.10\% & 10.80\% & 0.00\% \\
FCFS  & 98.35\% & 94.15\% & 36.45\% & 0.00\% \\
\bottomrule
\end{tabular}
\end{table}

Consider a request that is drafted at the edge and verified by the server within an iteration. Let $N^v$ denote the number of tokens accepted after verification in this iteration, and let $T_{\text{total}}$ denote the total time consumed by this iteration. The achieved token speed is:
\begin{equation}
    s = \frac{N^v}{T_{\text{total}}} 
    \label{eq:motivation:speed-def}
\end{equation}

By decomposing the iteration time as $T_{\text{total}} \;=\; T_{\text{draft}} \;+\; T_{\text{network}} \;+\; T_{\text{queue}} \;+\; T_{\text{verify}}$, the achieved token speed can be written as:
\begin{equation}
    s \;=\; \frac{N^v}{T_{\text{draft}} \;+\; T_{\text{network}} \;+\; T_{\text{queue}} \;+\; T_{\text{verify}}},
    \label{eq:motivation:time-breakdown}
\end{equation}
where $T_{\text{draft}}$ is the edge drafting time, $T_{\text{network}}$ is the communication time for sending the draft payload and receiving the verification outcome, $T_{\text{queue}}$ is the server-side waiting time before verification starts, and $T_{\text{verify}}$ is the verification execution time on the server. In our setting, $T_{\text{draft}}$ is determined by the edge device, and $T_{\text{network}}$ can be measured by timestamps; in contrast, $T_{\text{queue}}$ and $T_{\text{verify}}$ are dominated by server batching and interference.

We estimate $N^v$ using the draft length $N_d$ and the acceptance rate $\alpha \in [0,1]$:
\begin{equation}
    \hat{N}^v \;=\; \alpha \cdot N_d,
    \label{eq:motivation:nv-est}
\end{equation}
where $\alpha$ is determined by the selection of draft model, target model, and drafting strategy. Combining \eqref{eq:motivation:speed-def}--\eqref{eq:motivation:nv-est} yields an equivalent \emph{server-side budget} constraint:

\begin{equation}
    T_{\text{queue}} + T_{\text{verify}}
    \;\;\le\;\;
    \underbrace{\frac{\alpha \cdot N_d}{s_c} - T_{\text{draft}} - T_{\text{network}}}_{\tau_{c(i)}}.
    \label{eq:motivation:server-budget}
\end{equation}
which enforces the speed SLO of class $c$.

Here, $\tau_{c(i)}$ defines the per-request verification-side time budget required to achieve the target token speed. Importantly, \eqref{eq:motivation:server-budget} makes the core implication explicit: \emph{once it receives a verification request from a device, $T_{draft}$ and $T_{network}$ are known, satisfying the application-level token-speed objective reduces to controlling server-side queuing and verification latency}. This motivates why \acronym treats SLO-awareness as a design goal: the verification server must incorporate profiling and SLO-aware scheduling to explicitly manage $T_{\text{queue}}$ and $T_{\text{verify}}$, so that application-level token generation speeds can be met.

\subsection{Wasted Drafting Time}
\label{sec:motivations_wdt}
Distributed speculative LLM edge serving systems enable an edge device to collaboratively generate tokens with a more capable verification server hosting the target model. Following the speculative decoding procedure in Section.~\ref{sec:related_work:background}, in each iteration the edge device drafts a \emph{fixed} block of $K$ draft tokens $\mathbf{y}_{1:K} \triangleq (y_1,\dots,y_K)$ and submits them for verification. The verification server then checks the draft block against the target model in order, and the verification outcome can be summarized by the acceptance indicator $\{A_i\}_{i=1}^{K}$ and the induced stopping index $R$ and acceptance length $L$ defined in Eq.~\eqref{eq:spec:accept}--Eq.~\eqref{eq:spec:R-L}.

By construction, verification proceeds until the first rejection occurs at index $R$ (or $R=K+1$ if all tokens are accepted), hence only the verified prefix $\mathbf{y}_{1:L}$ with $L=R-1$ is committed to the response, while the remaining suffix $\mathbf{y}_{(L+1):K}$ is discarded. Consequently, when the edge drafts a fixed-length block, any drafted tokens beyond the realized acceptance length $L$ do not contribute to the final output of that iteration and therefore constitute wasted computation on the edge device.

\paragraph{Definition of WDT.}
When the edge drafts a fixed window $K$, any drafted token beyond the accepted prefix is \emph{over-drafted}. Following the standard notion of speculative ``waste,'' we define the number of wasted draft tokens as
\begin{equation}
W \triangleq (K - L)^+ = \max\{0, K-L\},
\end{equation}
which is exactly the portion of the fixed draft window that is rejected by the server. 
Let $\tau_d$ be the average per-token drafting time on the edge (or equivalently, $s_d = 1/\tau_d$ be the draft throughput in tokens/s). The \emph{WDT} in one speculate--verify iteration is defined as
\begin{equation}
T_{\mathrm{wdt}} \triangleq \tau_d \cdot W = \frac{W}{s_d}.
\end{equation}
Hence, the total edge drafting time can be decomposed into a useful part and a wasted part:
\begin{equation}
T_{\mathrm{draft}} = \tau_d K = \tau_d L + \tau_d (K-L)
= T_{\mathrm{draft}}^{\mathrm{use}} + T_{\mathrm{wdt}}.
\label{eq:draft_time_breakdown}
\end{equation}

\begin{figure*}[t!]
    \centering
    \includegraphics[scale=0.35]{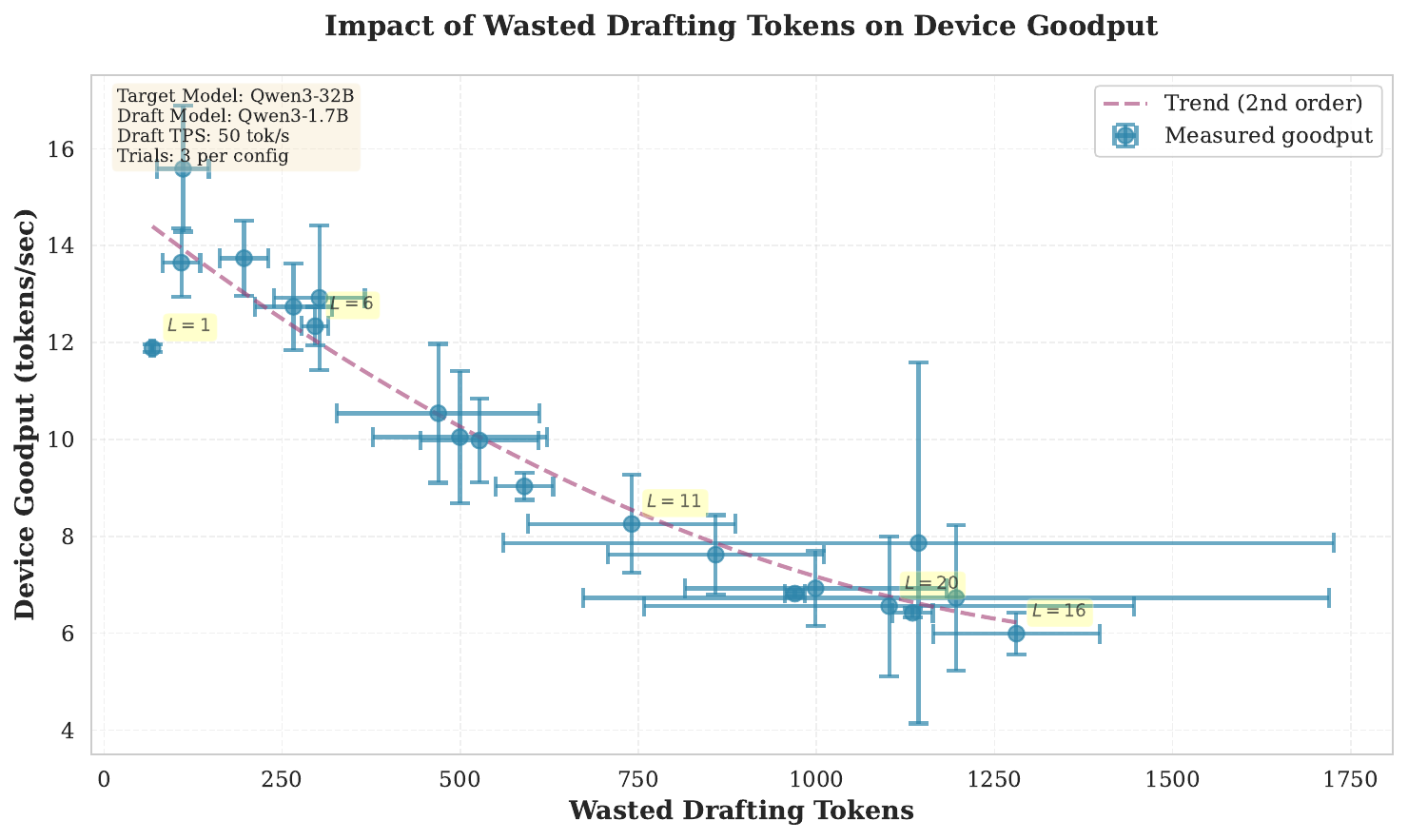}
    \caption{Impact of WDT on Device Goodput}
    \Description{Higher WDT leads to lower device goodput}
    \label{fig:wdt_goodput}
\end{figure*}

\paragraph{Impact on goodput and SLO}
\emph{Goodput} is the rate (tokens/s) of \emph{verified and committed} output tokens delivered to the application, excluding any draft tokens that are later rejected and discarded.
Figure.~\ref{fig:wdt_goodput} shows a clear negative correlation between wasted drafting tokens (a proxy for WDT) and the achieved device goodput: as waste increases, the device spends a growing fraction of its fixed drafting capacity (50~tokens/s) on tokens that never contribute to final output, causing the effective throughput to degrade. 

Moreover, by substituting $T_{draft}$ in Eq. \ref{eq:motivation:time-breakdown} with Eq. \ref{eq:draft_time_breakdown}, we have:
\begin{equation}
    s \;=\; \frac{N_v}{(T_{\mathrm{draft}}^{\mathrm{use}} + T_{\mathrm{wdt}}) \;+\; T_{\text{network}} \;+\; T_{\text{queue}} \;+\; T_{\text{verify}}},
    \label{eq:motivation:time-breakdown_wasted}
\end{equation}
which indicates that higher WDT leads to lower token generation speed, further leading to SLO violations.

\begin{figure*}[t]
  \centering
  \begin{subfigure}[t]{0.24\textwidth}
    \centering
    \includegraphics[width=\linewidth]{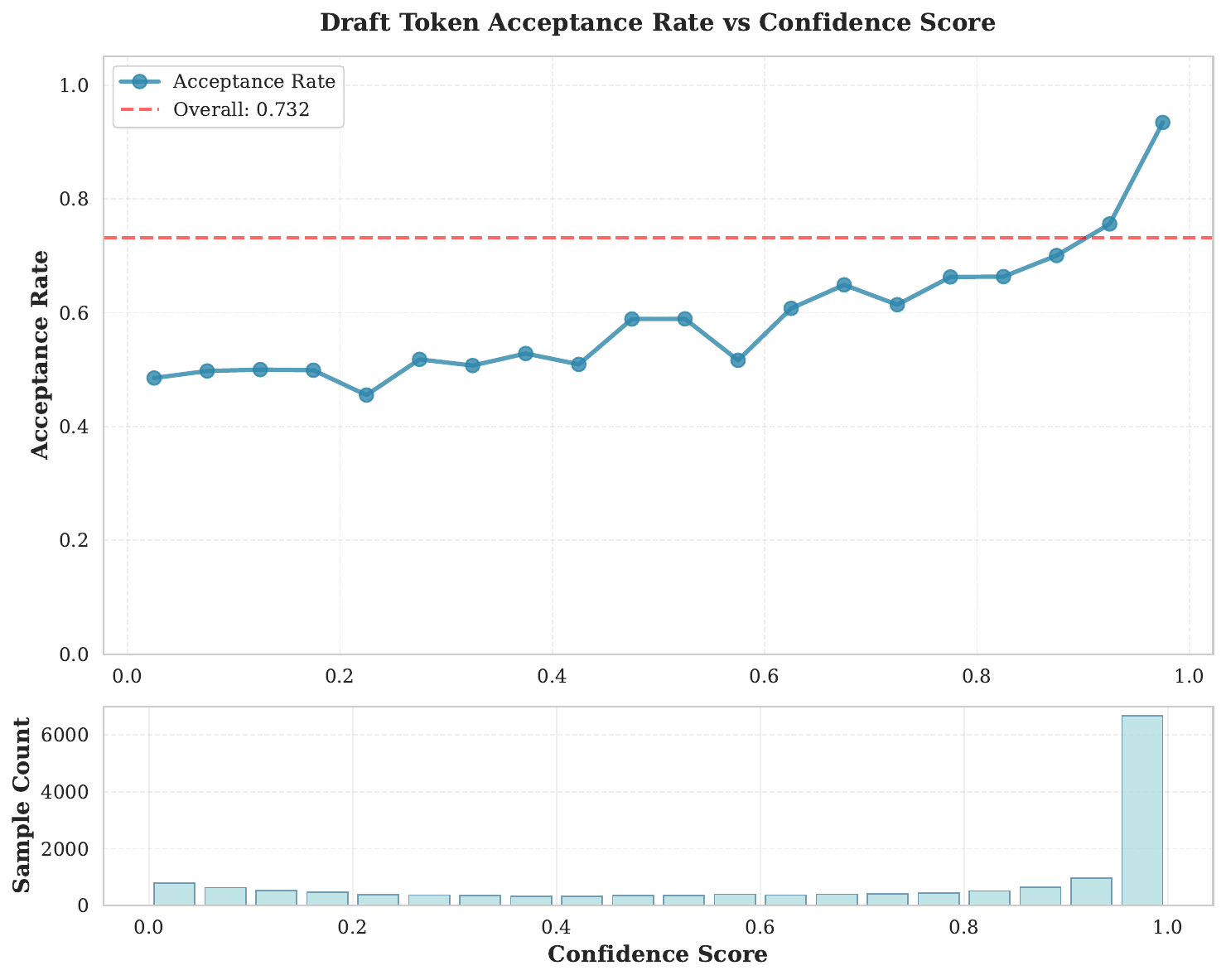}
    \caption{Confidence Score}
  \end{subfigure}\hfill
  \begin{subfigure}[t]{0.24\textwidth}
    \centering
    \includegraphics[width=\linewidth]{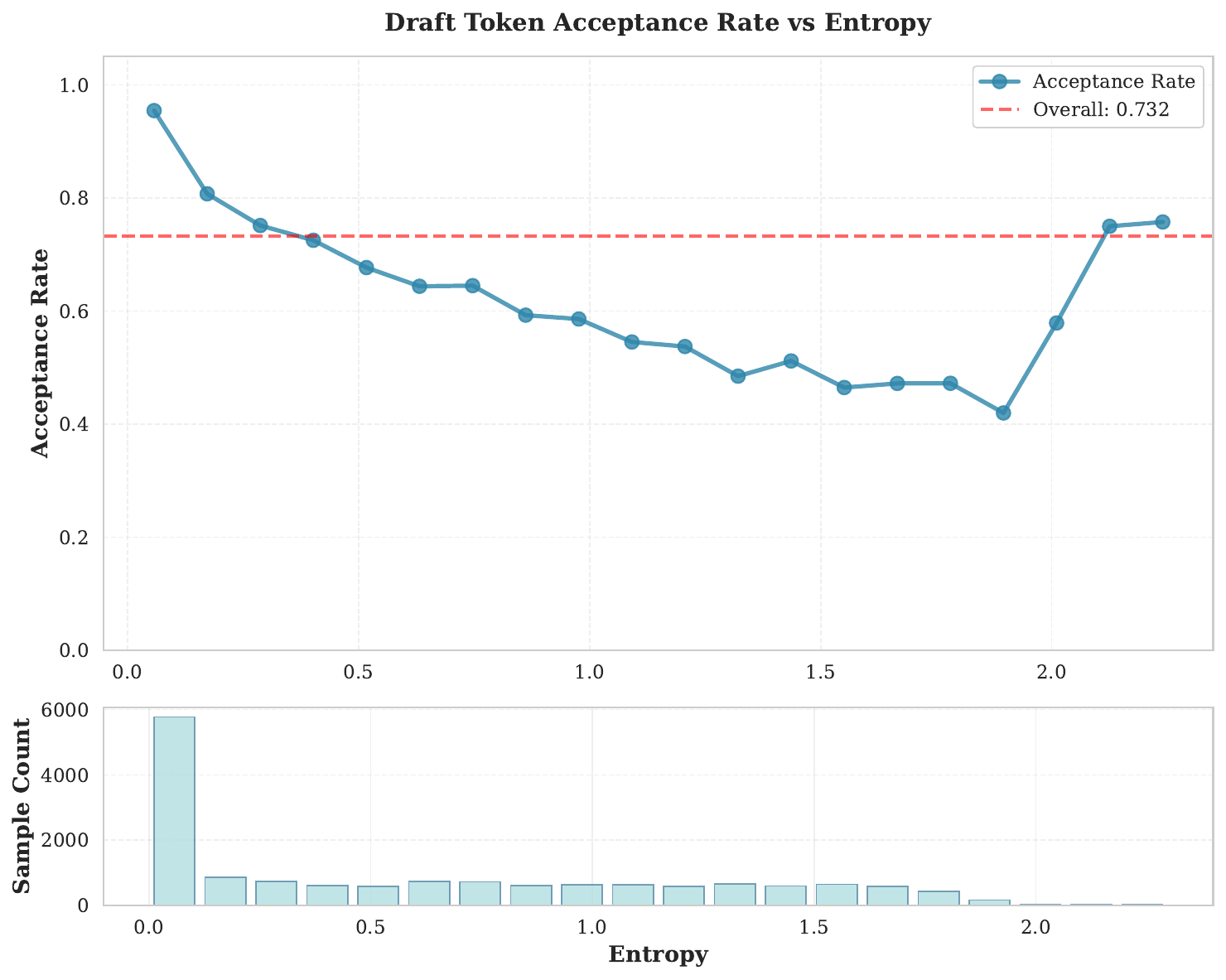}
    \caption{Entropy}
  \end{subfigure}\hfill
  \begin{subfigure}[t]{0.24\textwidth}
    \centering
    \includegraphics[width=\linewidth]{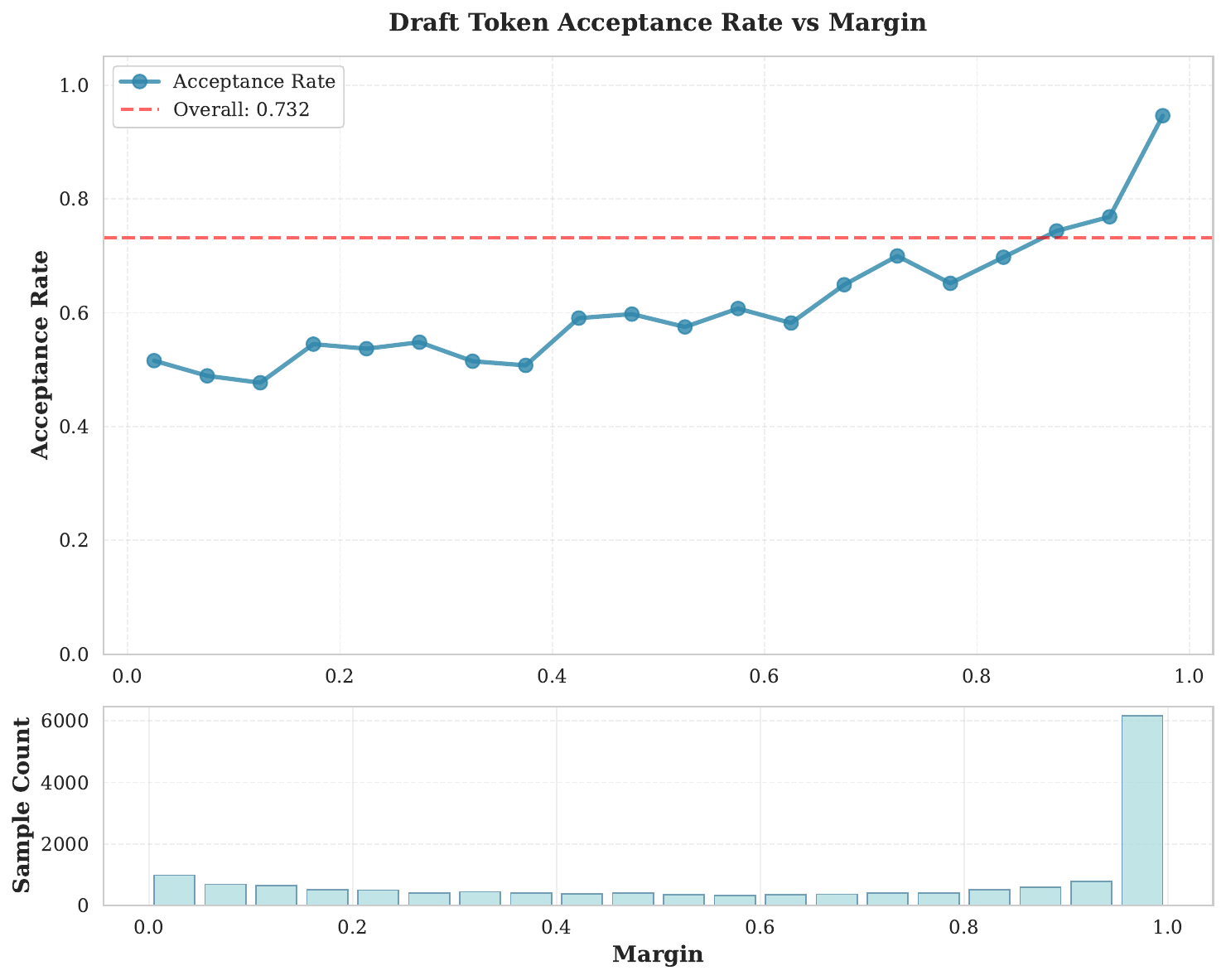}
    \caption{Margin}
  \end{subfigure}\hfill
  \begin{subfigure}[t]{0.24\textwidth}
    \centering
    \includegraphics[width=\linewidth]{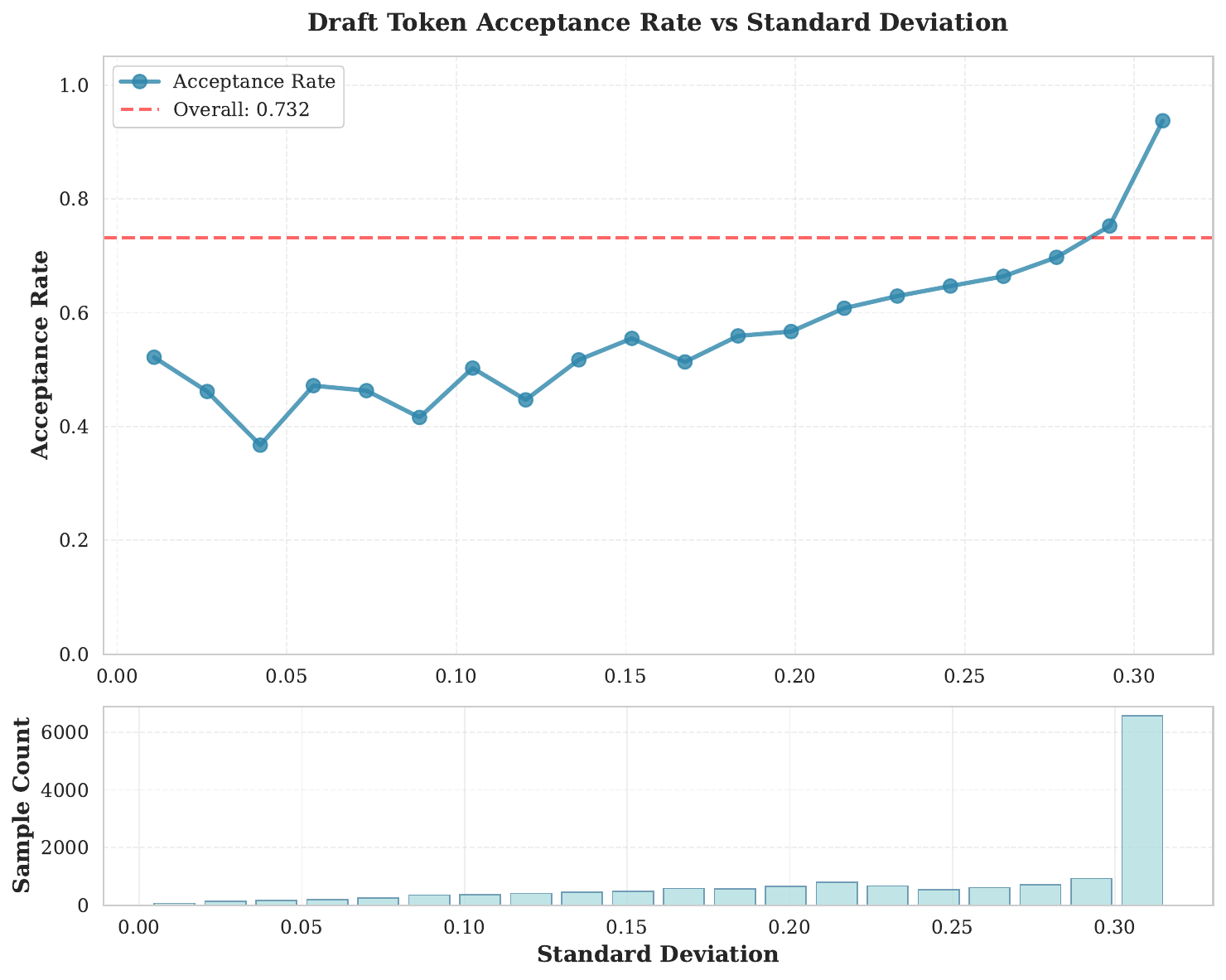}
    \caption{Standard Deviation}
  \end{subfigure}
  \caption{Relationships of Draft Logits Statistics with Acceptance Rate}
  \Description{The sub-figures show the clear impact of four statistics on the acceptance rate.}
  \label{fig:statistics_relationships}
\end{figure*}

\begin{figure*}[t]
    \centering
    \includegraphics[scale=0.2]{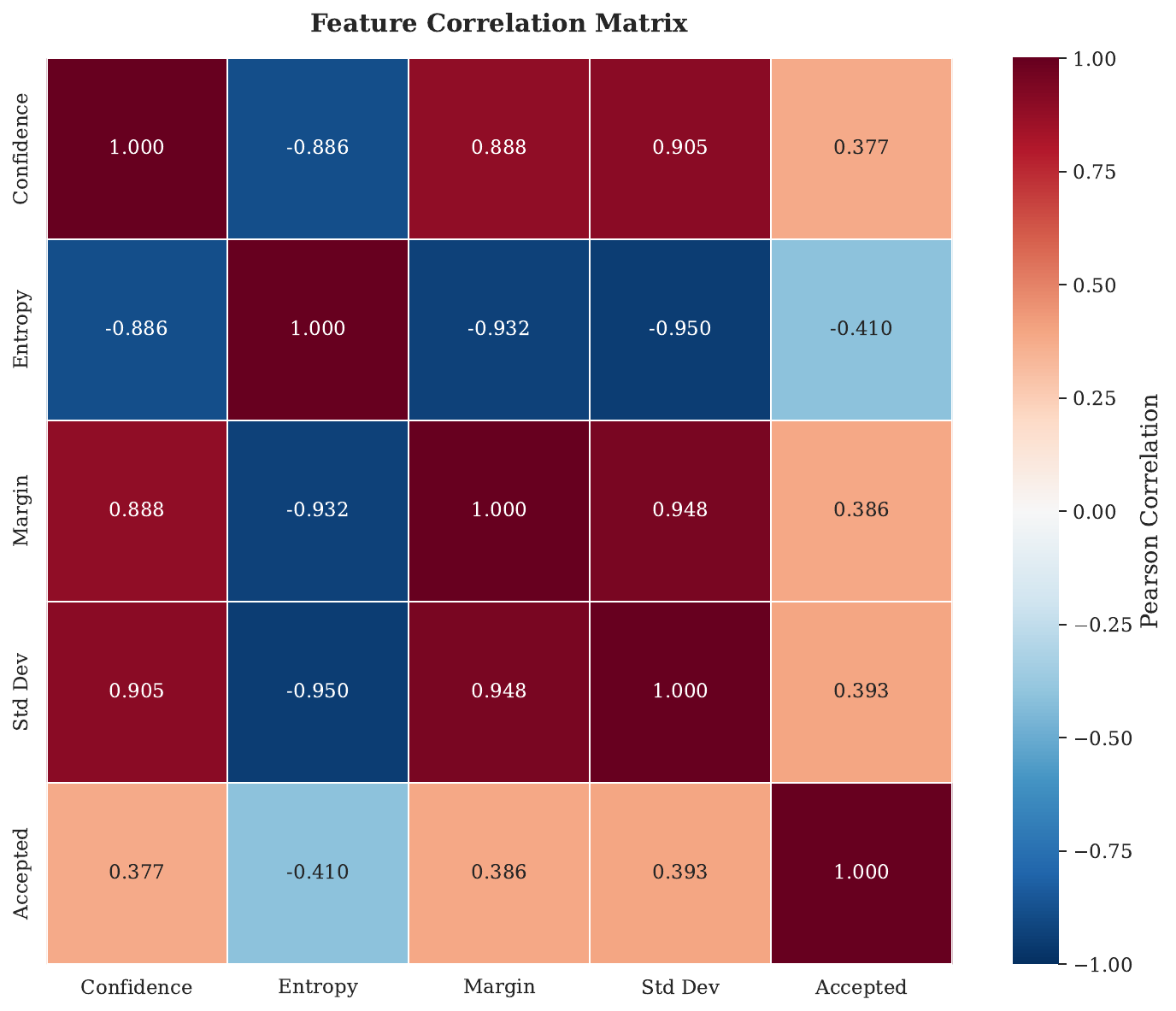}
    \caption{Correlation between Logits Statistics and Token acceptance}
    \Description{Confidence score, margin and standard deviation show }
    \label{fig:motivation_correlation_heatmap}
\end{figure*}

\subsection{Predictable Acceptance Rate}
Across a speculative session, draft tokens vary in difficulty: tokens requiring long-range context or domain knowledge are more likely to be rejected by the target model at verification time. Prior work \cite{huang2025specserve} and our observations show that this difficulty is reflected in draft-side logits. At step $t$, the draft model outputs a next-token distribution $p_t(\cdot)$ whose summary statistics (e.g., confidence, entropy, top-2 margin, and logit dispersion) quantify token-level uncertainty: a sharper distribution (higher confidence/larger margin/lower entropy) typically implies higher agreement with the target model and thus higher acceptance probability. We instrument \acronym to log these statistics together with verification outcomes and find a strong, consistent monotonic relationship (Fig.~\ref{fig:statistics_relationships}): acceptance increases with confidence/margin and decreases with entropy/dispersion, which is further supported by the Pearson correlation trends (Fig.~\ref{fig:motivation_correlation_heatmap}). Motivated by these signals, \acronym uses lightweight logit-derived features to predict imminent rejections online and trigger verification adaptively instead of drafting a fixed window, reducing WDT with negligible overhead.

Overall, WDT is a structurally unavoidable inefficiency in fixed-window speculative decoding: whenever the target model rejects early, the edge has already paid the full drafting cost for the unused suffix. This waste directly reduces verified token generation speed and tightens the SLO feasibility region, motivating \acronym’s design to explicitly control speculation so as to minimize WDT while preserving server efficiency.

\subsection{Verification Interference}
\label{sec:motivation:interference}
By batching tokens from multiple requests, the verification server consolidates many per-request launches into a few large GPU kernels, improving utilization and amortizing overheads \cite{kwon2023pagedattention,yu2022orca}. However, batching benefits are highly operator-dependent: GEMM-dominated QKV projections and Multilayer Perceptrons (MLPs) scale well with the aggregated token count, while self-attention is constrained by heterogeneous sequence lengths and KV-cache memory traffic (plus reductions such as softmax), yielding weaker and less uniform scaling \cite{vaswani2017attention,dao2022flashattention}. Consequently, practical serving systems often adopt \emph{selective batching}—aggressively batching linear operators while treating attention with more conservative batching and specialized kernels \cite{yu2022orca}.

\begin{figure*}[t!]
    \centering
    \includegraphics[scale=0.25]{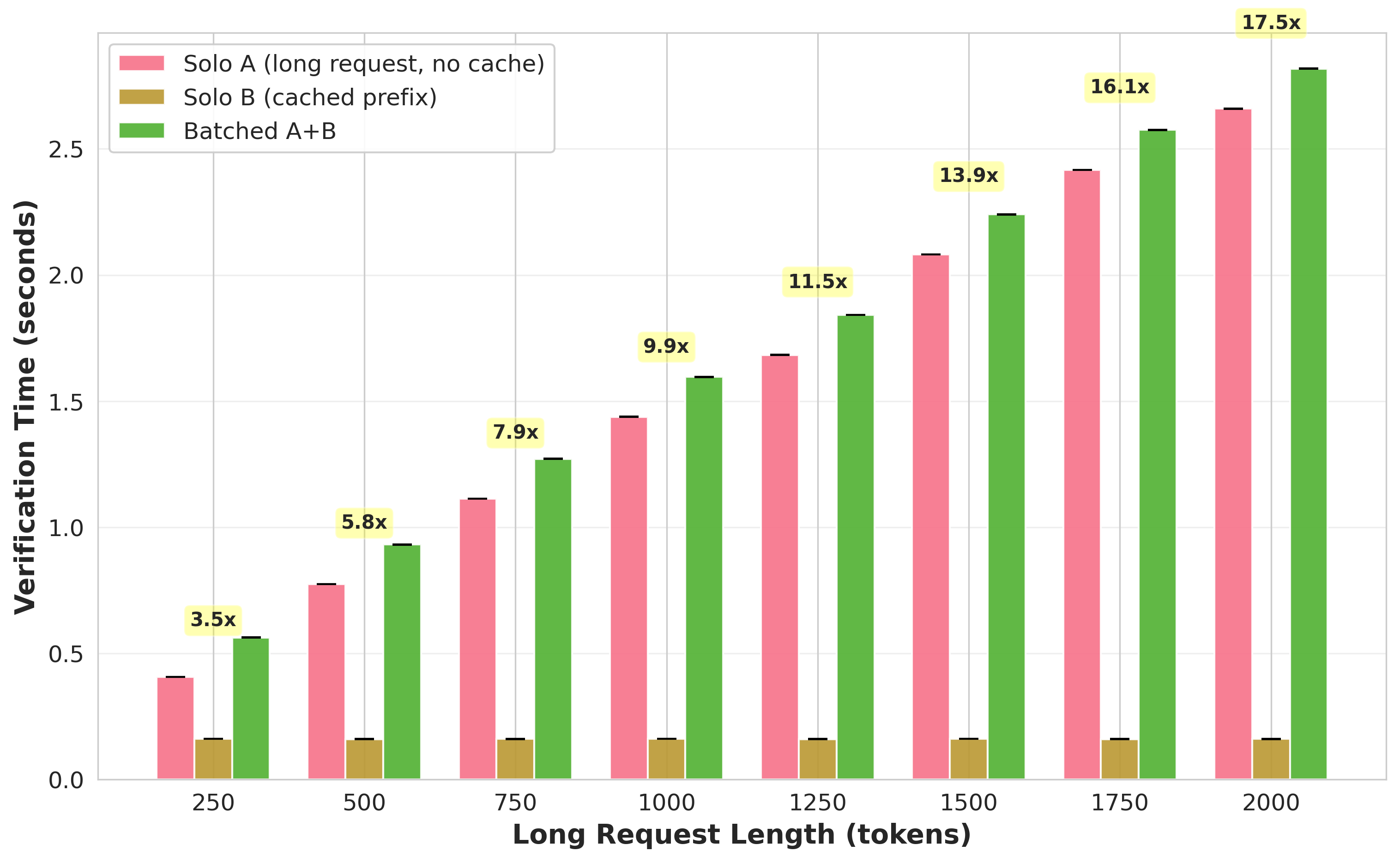}
    \caption{Verification Interference among Heterogeneous Requests}
    \Description{Long Uncached Verification Request Dominates the Verification Time in A Batch}
    \label{fig:motivation_verification_interference}
\end{figure*}

This operator heterogeneity (batch-friendly linear layers vs.\ less uniformly scalable attention) turns into a latency issue in speculative serving, where verification requests can arrive with different draft lengths. When such heterogeneous requests are co-batched to maximize GPU parallelism, the micro-batch becomes a synchronization barrier: it completes only after the slowest request finishes its layer stack. As a result, long verifications inflate the batch time and short verifications inherit the same completion time. We term this head-of-line effect \emph{verification interference}: a few long verification requests, which are typically bottlenecked by the less-batch-friendly attention/KV-cache path, can dominate the batch and increase the end-to-end latency of many short requests. Fig.~\ref{fig:motivation_verification_interference} is obtained by profiling vLLM verification latency of LLaMA-70B on two A100 GPUs, where the cached short request $B$ has a 500-token prefix plus 5 newly drafted tokens to verify. When executed in isolation, the long request (Solo A) increases with its verification length, whereas the cached short request (Solo B) remains nearly flat; however, once co-batched (Batched A+B), the batch completion time is dictated by the long request, causing the short request to suffer from the long-request latency.

\section{\acronym Design} \label{sec:sled}
Motivated by the lack of SLO-awareness,  the wasted drafting time, the predictable acceptance rate, and the Verification Interference observed in prior distributed speculative edge LLM serving systems, we present \acronym. In \acronym, user devices speculate draft tokens while the GPU cluster batches and verification requests, enabling the system to serve more edge devices with the same GPU capacity while maintaining high-quality, lossless LLM outputs. Beyond this baseline architecture, \acronym integrates an intelligent drafting controller, an SLO-aware scheduler, a verification time estimator, and a verification engine. Together, these components enable adaptive drafting and SLO-aware verification, improving drafting efficiency while reducing verification-side SLO violations under heterogeneous and time-varying workloads.

\textcolor{black}{To guarantee that \acronym\ remains strictly lossless, the verification engine follows the standard speculative sampling algorithm~\cite{leviathan2023fast}, as shown in Section~\ref{sec:related_work:background}. According to the mathematical proof in Leviathan's work~\cite{leviathan2023fast}, the verification role guarantees that the resulting output distribution is identical to that produced by decoding from the target model alone. Importantly, physically separating the drafter (edge) and verifier (cloud) in \acronym\ only affects the latency of verification, but does not change the token distributions involved or the acceptance/rejection logic itself, hence the output quality of \acronym is lossless. The output qualities of \acronym and traditional auto-regressive decoding are tested and compared in Appendix~\ref{sec:appendix:quality}.}

On the verification server, \acronym further implements PageAttention \cite{kwon2023pagedattention} and a prefix cache \cite{gim2024promptcache} to reuse KV states across consecutive verification rounds for the same response, thereby increasing verification throughput and mitigating redundant verification work. To the best of our knowledge, \acronym is the first comprehensive distributed speculative edge LLM serving system that jointly provides dynamic drafting, SLO-aware control and interference-aware verification batching (i.e., explicitly addressing Verification Interference), alongside practical cache optimizations on the verification server.

\begin{figure*}[t!]
    \centering
    \includegraphics[scale=0.3]{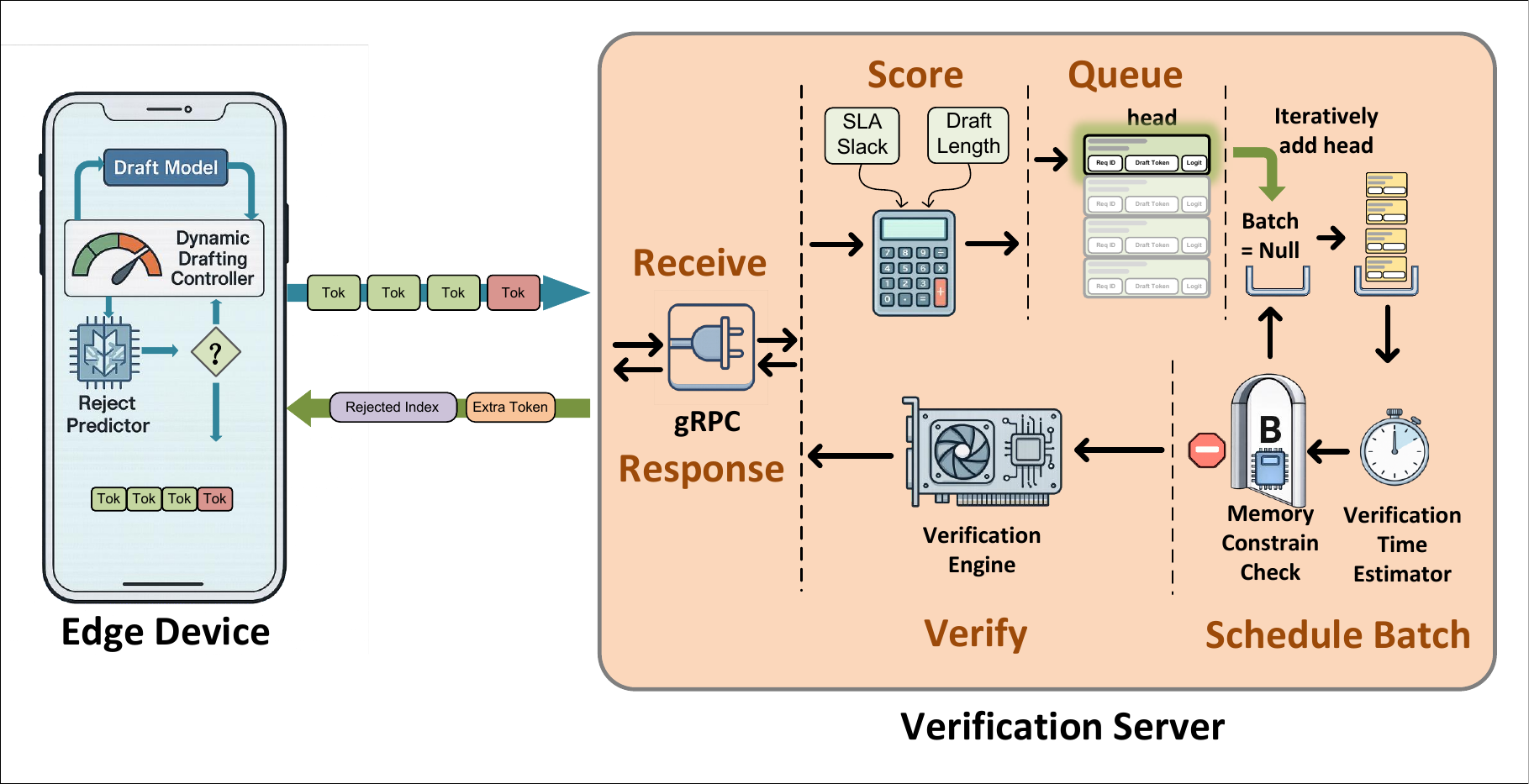}
    \caption{\acronym architecture and data flow}
    \Description{Detailed architecture and data flow in \acronym system: }
    \label{fig_data_flow_overview}
\end{figure*}

Figure. \ref{fig_data_flow_overview} shows the basic architecture of \acronym and data flow among edge device and verification server. On edge devices, a small draft model is selected according to the device's memory and computation constraint and then deployed. The intelligent drafting controller samples candidate token from draft model and verify it with a light-weight predictor for confidence checking iteratively, until the predictor signals a verification request is needed. Then, draft tokens are transmitted to verification server through gRPC \cite{grpcgithub} call. Once request arrives at verification server, it will be sent to either \emph{utility queue} or \emph{urgent queue} according to the request's SLO slack and the utility score that reflects the expected goodput contributed by this request. Meanwhile, an asynchronized batch scheduler collect a certain number of request from the queues to form a verification batch without violating the memory constrain and SLO deadline. Thereafter, the batch will be forwarded through verification engine to get the verification results, namely rejected position and extra token, which will then be responded to edge device.

\subsection{Intelligent Drafting Controller}
In the distributed speculative edge serving systems like \acronymsled, a fixed number of draft tokens are generated and sent to verification server. As motivated in Section.\ref{sec:motivations_wdt}, we observe the strong relationship between acceptance possibility with the statistics of draft logits, which can be used to prevent over-drafting. Therefore, we propose an intelligent drafting controller over the draft model to adjust speculative length dynamically. 

Specifically, a machine learning based predictor is designed to give the prediction of whether this draft token will be accepted by target model according to the input features of: confidence score, the entropy, the margin, and the standard deviation. Based on the prediction, we adopt a simple stop-at-first-predicted-rejection strategy to request for verification from the verification server. We analyze the relationship of predictor and the WDT, and have the Theorem~\ref{thm:wdt} below:

\begin{theorem}[Monotonicity of WDT]
With predictor model $\theta'$ trained with lower false alarm rate than model $\theta$, we have:
\begin{equation}
    \mathbb{E}[W_{\theta'}] \le \mathbb{E}[W_{\theta}],
\end{equation}
\label{thm:wdt}
\end{theorem}

Theorem~\ref{thm:wdt} shows that, by training a predictor with lower false alarm rate, we are able to reduce the WDT, hence improving system goodput and drafting-verification round between user devices and verification server.

\paragraph{Rejection predictor design.}
Inspired by Theorem~\ref{thm:wdt}, we design a lightweight binary \emph{rejection predictor} that estimates whether the \emph{next} drafted position is likely to be the first target-side rejection within the current speculative window, enabling a stop-at-first-predicted-rejection policy that triggers verification early and avoids over-drafting. The predictor operates purely on draft-side signals: for each token position, we compute a compact feature vector from summary statistics of the draft logits, and train a small edge-friendly classifier to distinguish \emph{accepted} tokens from the \emph{first rejected} token; positions after the first rejection are intentionally ignored since verification terminates immediately and those suffix tokens do not affect either the acceptance length or wasted work.

\subsection{Verification Batch Scheduling Problem}
With verification requests from edge devices, \acronym dynamically batches the verification requests for execution, ensuring higher GPU utilization and system throughput. Meanwhile, verification requests with diverse application types need to be served to fulfill the required token generation speed. In \acronym, We model the batch scheduling problem as a knapsack problem and then propose an SLO-aware scheduling algorithm.

According to the server-side processing budget in Eq.\ref{eq:motivation:server-budget}, we model the verification SLO for request $i$ arrives at time $a_i$ as a relative deadline:
\begin{equation}
d_i \triangleq a_i + \tau_{c(i)},
\qquad
\end{equation}

A request meets its verification SLO if its completion time $C_i$ satisfies $C_i \le d_i$.
This deadline-based SLO abstraction is standard in real-time scheduling and deadline-aware serving systems
~\cite{liu1973edf,zhang2023bcedge}.


We formulate verification batching as an SLO-constrained long-run goodput maximization problem. Let $t_k$ denote the $k$-th dispatch epoch, at which the scheduler selects a batch $B_k$ from the pending requests and submits it to the GPU. Each request $i$ carries an estimated number of verified tokens $\hat{N}^v_i$ , defined in Eq. \ref{eq:motivation:nv-est}, an estimated memory footprint $\hat{m}_i$, and a deadline $d_i$ derived from its class $c$. The objective maximizes the time-average total goodput accumulated by all dispatched batches over an infinite horizon. Each batch must satisfy a GPU memory constraint, and an SLO feasibility constraint requiring every included request to complete before its deadline. The optimization problem is shown in Eq. \ref{eq:hard_slo_obj}-\ref{eq:optimization_obj_2}.
\begin{align}
\max_{\{B_k\}} \quad &
\liminf_{T\to\infty}\frac{1}{T}\,
\mathbb{E}\!\left[
\sum_{k:\, t_k < T}
\sum_{i\in B_k}
\hat{N}^v_i
\right]
\label{eq:hard_slo_obj}\\
\text{s.t.}\quad &
\sum_{i\in B_k} \hat{m}_i \le M(t_k), \quad \forall k, \label{eq:optimization_obj_1} \\
&
t_k + \hat{T}_{batch}(B_k) \le d_i,\quad \forall i\in B_k,\ \forall k.
\label{eq:optimization_obj_2}
\end{align}

The optimization problem captures the common ``efficiency under SLO'' goal, which is widely adopted in production LLM inference scheduling~\cite{bari2025slai,huang2025multislo}. By solving optimization problem in Eq. Eq. \ref{eq:hard_slo_obj}-\ref{eq:optimization_obj_2}, system goodput can be maximized while SLO requirements being met. However, even with a simplified runtime model, selecting $B_k$ under memory and deadline constraints is NP-hard. Moreover, $\hat{T}_{batch}(B_k)$ is workload-dependent (batch size and token-length heterogeneity), making the exact per-batch optimization intractable, which motivates lightweight online heuristics, which has been adopted in prior SLO-aware inference schedulers~\cite{seo2021slo,bari2025slai}. 

To derive a practical scheduler, we adopt a standard per-epoch reduction: at each dispatch time $t_k$, we treat the available GPU budget as a bounded resource and seek a \emph{locally} optimal batch that delivers the largest incremental goodput while remaining SLO-feasible. Specifically, we approximate the marginal verification goodput of admitting a request by a profiling the number of its expected verified tokens and its marginal verification cost. Under this view, the per-epoch batch selection becomes a knapsack-like packing problem: each request $i$ provides value $\hat{g}_i$ (expected accepted tokens) and consumes verification cost $\hat{v}_i$, and we aim to pack high-value items within the feasible budget. This motivates our utility score as a \emph{value density}:
\begin{equation}
U_i \triangleq \frac{\hat{N}^v_i}{\hat{v}_i},
\end{equation}
which measures expected goodput gained by admitting verification request $i$ in the batch. 

\subsection{Scheduling Algorithm}

In this work, we propose a greedy algorithm to select batches from decreasing $U_i$, which is a principled approximation to the per-epoch knapsack objective. Meanwhile, in order to guarantee the SLO, we deliberately keep SLO as a \emph{constraint/priority override}. Precisely, We introduce a \emph{time-critical fast path} that overrides density ordering, following the common ``critical-first, best-effort-later'' pattern in SLO-aware inference schedulers~\cite{seo2021slo,bari2025slai}.

For each request $i$, we define a conservative \emph{latest start time} (LST):
\begin{equation}
\mathrm{LST}_i \triangleq d_i - \hat{v}_i - \delta,
\label{eq:lst}
\end{equation}
where $\delta$ is a guard time for modeling error and queueing uncertainty, a standard technique in SLO-aware serving to absorb estimation errors and tail effects, and it's determined from experience.

A request becomes \emph{urgent} and put into urgent queue at time $t$ if $t \ge \mathrm{LST}_i$.
Urgent requests are ordered by earliest deadline (EDF), while non-critical requests are ordered by utility score $U_i$, forming the utility queue.

At each dispatch epoch $t_k$, the planner builds a batch incrementally:
it first admits EDF-ordered critical requests,
and then fills the residual capacity using utility score-ordered requests.
While requests being added to batch repeatedly, each tentative batch is validated by both constraints (Eq. \ref{eq:optimization_obj_1}- \ref{eq:optimization_obj_2}), until any violation is seen. The scheduling algorithm is shown in Algorithm. \ref{alg:sled2-batch-scheduler}.

\begin{algorithm}[t]
\scriptsize
\caption{SLO-aware Greedy Verification Batch Scheduling at Dispatch Epoch $t_k$}
\label{alg:sled2-batch-scheduler}
\begin{algorithmic}[1]
\Require Pending request set $\mathcal{Q}(t_k)$; memory budget $M(t_k)$; guard time $\delta$;
         per-request estimates $\{\hat{g}_i,\hat{v}_i,\hat{m}_i,d_i\}_{i\in\mathcal{Q}(t_k)}$;
         batch-time estimator $\hat{T}(\cdot)$.
\Ensure Dispatched batch $B_k$.

\Function{FeasibleAdd}{$B, i, t_k$}
    \State $B' \gets B \cup \{i\}$
    \State $m(B') \gets \sum_{j \in B'} \hat{m}_j$
    \State $D_{\min}(B') \gets \min_{j \in B'} d_j$
    \State \Return $\big(m(B') \le M(t_k)\big)\ \land\ \big(t_k + \hat{T}(B') \le D_{\min}(B')\big)$
\EndFunction

\Statex
\ForAll{$i \in \mathcal{Q}(t_k)$}
    \State $U_i \gets \hat{g}_i / \hat{v}_i$ \Comment{Utility (value density)}
    \State $\mathrm{LST}_i \gets d_i - \hat{v}_i - \delta$ \Comment{Latest start time (Eq.~\ref{eq:lst})}
\EndFor

\State $\mathcal{Q}_{\mathrm{crit}} \gets \{ i \in \mathcal{Q}(t_k) \mid t_k \ge \mathrm{LST}_i \}$
\State $\mathcal{Q}_{\mathrm{non}} \gets \mathcal{Q}(t_k) \setminus \mathcal{Q}_{\mathrm{crit}}$
\State Sort $\mathcal{Q}_{\mathrm{crit}}$ by increasing $d_i$ \Comment{EDF among critical requests}
\State Sort $\mathcal{Q}_{\mathrm{non}}$ by decreasing $U_i$ \Comment{Greedy by utility score}

\State $B \gets \emptyset$
\State $\textit{stop} \gets \textbf{false}$

\Comment{(1) Time-critical fast path: admit EDF-ordered critical requests first}
\ForAll{$i \in \mathcal{Q}_{\mathrm{crit}}$}
    \If{\Call{FeasibleAdd}{$B,i,t_k$}}
        \State $B \gets B \cup \{i\}$
    \Else
        \State $\textit{stop} \gets \textbf{true}$ \Comment{Violation observed; stop growing batch}
        \State \textbf{break}
    \EndIf
\EndFor

\Comment{(2) Best-effort fill: only if all critical admissions did not trigger violation}
\If{\textbf{not} $\textit{stop}$}
    \ForAll{$i \in \mathcal{Q}_{\mathrm{non}}$}
        \If{\Call{FeasibleAdd}{$B,i,t_k$}}
            \State $B \gets B \cup \{i\}$
        \Else
            \State \textbf{break} \Comment{Violation observed; stop growing batch}
        \EndIf
    \EndFor
\EndIf

\State $B_k \gets B$
\State \Return $B_k$ \Comment{Dispatch $B_k$ at time $t_k$; remaining requests stay pending}
\end{algorithmic}
\end{algorithm}

\subsection{\textcolor{black}{Verification Time Estimator}}

To enforce SLO constraints while dynamically constructing micro-batches, we require a lightweight estimator that predicts verification latency $T_{\text{batch}}$. A critical challenge in speculative serving is that a single micro-batch often contains a heterogeneous mix of verification requests in different verifying stages. We distinguish between two primary verification types that coexist in the batch:
\begin{itemize}

    \item \textbf{First-time Verification (Cold Start):} The request is new to the server (or its cache was evicted). The server must process the entire prompt sequence plus the initial draft tokens. Here, the number of new tokens $L_{\text{new}}$ is large (typically hundreds), while the cached length $L_{\text{cached}}$ is zero.
    \item \textbf{Follow-up Verification (Warm Cache):} The request has valid history in the KV cache. The server only processes the small set of newly generated draft tokens (e.g., 3--5 tokens). Here, $L_{\text{new}}$ is small, but the draft tokens must attend to a long history $L_{\text{cached}}$.
\end{itemize}

Building on the operator heterogeneity analysis in Section~\ref{sec:motivation:interference}, we model the total batch latency as an additive combination of three orthogonal components that capture these diverse states: linear computation, attention interaction, and memory access.

\paragraph{Computational modeling.}
First, we analyze the linear operations (embeddings, QKV projections, FFNs). These layers are stateless and process each token independently. Regardless of whether a request is a "first-time" prompt or a "follow-up" draft, the computational cost scales strictly with the number of tokens entering the GPU. Thus, the aggregate linear complexity is simply the sum of new tokens across all requests in the batch: $N_{\text{linear}} = \sum_i L_{\text{new},i}$.

Second, we analyze the attention mechanism, which is highly sensitive to the verification type. For a \textit{first-time} verification, the new tokens attend primarily to each other (quadratic self-attention). For a \textit{follow-up} verification, the few draft tokens must primarily attend to the long cached history (linear cross-attention). We unify these behaviors under a single complexity metric. For every request $i$, the query vectors of the $L_{\text{new},i}$ tokens must compute dot products with keys from the entire sequence context $L_{\text{total},i} = L_{\text{cached},i} + L_{\text{new},i}$. We define the aggregate attention complexity $N_{\text{interactions}}$ as the total count of query-key interactions:
\begin{equation}
N_{\text{interactions}} = \sum_{i} (L_{\text{total},i} \times L_{\text{new},i})
\end{equation}
This formulation automatically adapts to the request type: it scales quadratically for cold starts (where $L_{\text{total}} \approx L_{\text{new}}$) and linearly with prefix length for warm follow-ups.

Finally, to capture the memory-bandwidth bottleneck, we quantify the IO overhead by the total count of cached tokens that must be loaded from HBM to the streaming multiprocessors: $N_{\text{cached}} = \sum_i L_{\text{cached},i}$.
Combining these factors, we estimate the batch verification latency as:
\begin{equation}
\hat{T}_{\text{batch}} = a \cdot N_{\text{linear}} + b_{\text{compute}} \cdot N_{\text{interactions}} + b_{\text{read}} \cdot N_{\text{cached}} + c
\label{eq:batch_time}
\end{equation}
where $c$ represents constant system overheads (e.g., kernel launches).

\paragraph{Profiling and validation.}
We fit this model using Ordinary Least Squares (OLS) on a dataset profiled on NVIDIA A100 GPUs with \textsc{Qwen3-32B}, ensuring coverage of compute-bound, memory-bound, and mixed regimes. The latency estimator achieves high accuracy with $R^2 > 0.99$ and a Mean Absolute Error (MAE) of $3.38$ ms on the test set. In the online system, we apply a small guard time $\delta$ to this prediction to absorb residual worst-case errors. \textcolor{black}{These coefficients are deployment-specific rather than per-request dynamic: once profiled for a fixed GPU type, target model, quantization setting, and inference backend, the same latency model can typically be reused during normal operation. Re-profiling and coefficient refitting are only needed when the execution stack changes materially, e.g., moving to a different GPU platform, switching kernels/backends, or changing the target model family/size or quantization setup.}

\textcolor{black}{\paragraph{Robustness to estimator error.}
WISP is designed to remain robust under moderate latency-estimation error. Let the true batch latency be
\begin{equation}
\label{eq:batch_latency}
    T_{\mathrm{batch}}(B)=\hat{T}_{\mathrm{batch}}(B)+\varepsilon(B),
\end{equation}
where $\varepsilon(B)$ is the prediction error. In our scheduler, the guard time $\delta$ in the latest-start-time rule serves as an explicit robustness margin against such error. Therefore, moderate underestimation mainly reduces the available SLO slack, while overestimation mainly makes the scheduler more conservative and lowers utilization or goodput. In other words, the effect is graceful rather than catastrophic: estimator error primarily shifts the slack--utilization tradeoff, rather than breaking correctness or the overall scheduling logic.}


\subsection{Verification Engine}
\label{sec:design:verification-engine}

\begin{figure*}[t!]
    \centering
    \includegraphics[scale=0.3]{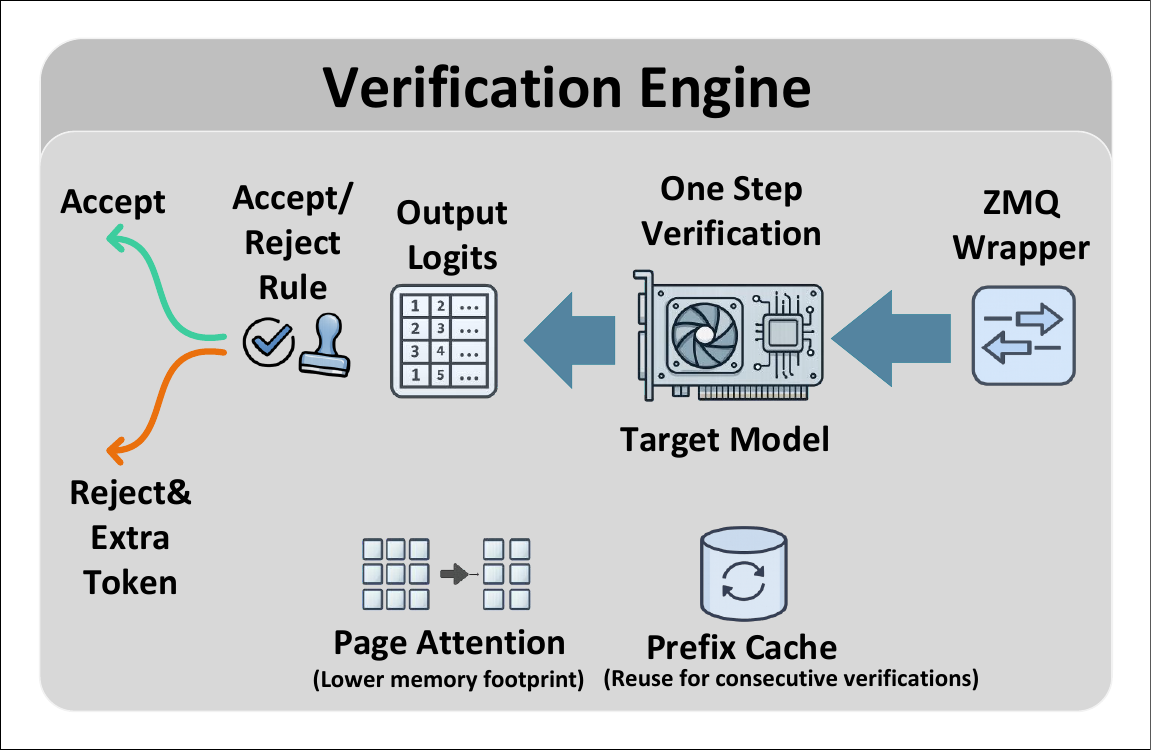}
    \caption{Architecture of verification engine}
    \Description{Detailed architecture and data flow of verification engine: }
    \label{fig:sled2_verification_engine}
\end{figure*}

After the SLO-aware scheduler assembles a verification batch, \acronym executes it in the \emph{verification engine}, which is responsible for (i) forwarding the batched inputs through the target model, (ii) applying the speculative accept/reject rule to obtain the rejected position and the \emph{extra token}, and (iii) maintaining per-request KV states to accelerate \emph{consecutive verifications} for the same response.
Figure.~\ref{fig:sled2_verification_engine} illustrates the main pipeline.

\paragraph{Batched one-step verification and speculative outputs.}
Given a verification batch $B_k$, the target engine performs \emph{one-step verification} by forwarding each request's newly drafted block $\mathbf{y}$ on top of its committed-prefix KV states. Concretely, it reuses the cached KV for the already accepted prefix and runs a prefill/extend pass only for the $K$ new draft tokens, producing logits $\{\boldsymbol{\ell}_1,\dots,\boldsymbol{\ell}_K\}$ (and the next-token logits when needed). For each request, it then applies the standard in-order accept/reject rule and stops at the first mismatch \cite{leviathan2023fast}: letting $R$ be the first rejection index (or $R{=}K{+}1$ if none) and $L{=}R{-}1$ the accepted length, the engine returns $(R,L)$ and an \emph{extra token} $\tilde{y}$ sampled from the target model at the rejection position to preserve the correct output distribution.

\paragraph{Efficient consecutive verifications with PageAttention and prefix cache.}
Distributed speculative serving triggers many consecutive speculate--verify rounds per response, so verification requests repeatedly share a long, growing prefix; re-verifying from scratch would redundantly recompute prefix prefill and waste GPU time/bandwidth. \acronym avoids this with two complementary engine optimizations: (i) a PageAttention-style paged KV layout that stores per-request KV in fixed-size blocks to reduce fragmentation and support dynamic sequence growth under tight GPU memory budgets \cite{kwon2023pagedattention}, and (ii) a per-session \emph{prefix cache} that persists committed-prefix KV across rounds so the target model forwards only the newly drafted block and updates KV \emph{in-place} \cite{gim2024promptcache}. Together with batched one-step verification, these mechanisms keep verification \emph{compute-efficient} (large coalesced GPU kernels) and \emph{state-efficient} (minimal redundant prefill), enlarging the feasibility region for meeting token-speed SLOs under heterogeneous, interference-prone workloads.

\section{Evaluation}
\label{sec:experiments}
We evaluate \acronym with the goal of answering three questions: (i) \textbf{system-level impact}---how much \acronym improves \emph{system capacity} (max supported edge devices under token-speed SLOs) and \emph{system goodput} compared to prior distributed speculative serving and centralized serving; (ii) \textbf{SLO behavior under contention}---whether SLO-aware batching can reduce token-speed violations under heterogeneous SLO classes and fluctuating verification workloads; and (iii) \textbf{component-level fidelity}---whether the intelligent drafting controller and the verification-time estimator provide the necessary accuracy/overhead trade-offs to be used in the online control loop. 

\subsection{Settings}
\paragraph{Experimental setup.}
We compare \acronym against (i) \textbf{centralized serving}, where all generation runs on the GPU server, and (ii) \textbf{SLED}, a distributed speculative baseline with FCFS verification; The verification server is implemented with vLLM using \emph{PageAttention} and a \emph{prefix cache} to reuse KV states across consecutive verifications, and edge--server communication uses gRPC. 

\textcolor{black}{To comprehensively validate the \acronym and conduct careful comparisons with baselines in end-to-end evaluations, we deploy the target LLMs, including Qwen3-32B, Qwen3-14B and Llama3.1-70B-Instruct, on an NVIDIA A100 80GB GPU, while running the lightweight draft models, including Qwen3-0.6B/1.7B/4B/8B, on edge devices.}
We measure predictor overhead on Raspberry Pi 5/4B CPUs. We generate load using a scalable verification-workload simulator that emulates many concurrent devices and produces verification requests across speculate--verify iterations, with four token-speed SLO classes $\{2,4,6,8\}$~tok/s; we log per-iteration drafting, network, queueing, and verification times to compute achieved token speed, and mark an SLO violation when the achieved speed falls below the class target.

\subsection{System Capacity and Throughput Analysis}
We evaluate \acronym at the system level by quantifying (i) \emph{system capacity}---the maximum number of concurrent edge instances supportable under token-speed SLOs---and (ii) \emph{system goodput}---the aggregate rate of \emph{verified and committed} tokens. We fix the verifier to a single NVIDIA A100 80GB and use a scalable verification-workload simulator to emulate $N$ edge devices that iteratively draft tokens and send verification requests. Each device is assigned a token-speed requirement $c \in \{2,4,6,8\}$~tokens/s, and we compute all metrics from per-batch server logs.

\subsubsection{System capacity}
For each service level $c$, we define capacity as the largest $N$ whose measured SLO violation rate stays below a threshold $\epsilon$:
\begin{equation}
\mathrm{Cap}(c) \triangleq \max \left\{ N ~\middle|~ \Pr\!\left[\text{token-speed} < c\right] \le \epsilon \right\}.
\end{equation}
We sweep $N$ upward, run the system to steady state, and estimate $\Pr[\cdot]$ from end-to-end token-speed traces (including queueing and verification time). Table~\ref{tab:system_capacity} summarizes the results for \acronym, \acronymsled, and centralized serving.

\acronym achieves the highest system capacity across all SLO classes $c$ and both target models, with particularly strong gains under tighter SLOs (larger $c$). \textcolor{black}{For Qwen3-32B, \acronym scales to 164/142/103/86 devices across Classes 1--4, compared to 97/80/54/41 for centralized serving and 83/42/27/21 for \acronymsled, translating to $1.69$--$2.10\times$ speedup over centralized and $1.98$--$4.10\times$ over SLED; for Qwen3-14B and Llama3.1-70B, the speedups are also profound.} The widening gap at tight SLOs is explained by two orthogonal controls: dynamic speculation and an SLO-aware batch scheduler, together with \acronym's prefix-cache reuse. In contrast, \acronymsled does not use caching and must recompute prefixes on every verification, so verification compute and interference escalate quickly as the SLO tightens; meanwhile, when the supported device count is already large (looser SLOs / higher aggregate throughput), fixed per-batch overheads (e.g., dispatch/scheduling and kernel-launch effects) occupy a larger fraction of time, making the multiplicative speedup comparatively smaller, consistent with the more modest gains in the highest-capacity regimes.

\newcommand{\black}[1]{\textcolor{black}{#1}}

\begin{table}[t]
\centering
\scriptsize
\setlength{\tabcolsep}{6pt}
\renewcommand{\arraystretch}{1.15}
\caption{System capacity (max supported devices) under different SLO classes. Higher is better.}
\label{tab:system_capacity}
\begin{tabular}{llcccc}
\toprule
\textbf{Systems} & \textbf{Target Model} &
\textbf{Class 1} & \textbf{Class 2} & \textbf{Class 3} & \textbf{Class 4} \\
\midrule
\multirow{2}{*}{\acronymsled}
& Qwen3-32B & $83$ & $42$ & $27$ & $21$ \\
& Qwen3-14B & $119$ & $88$ & $71$ & $52$ \\
& \black{Llama-70B} & \black{$48$} & \black{$17$} & \black{$7$} & \black{$3$} \\
\midrule
\multirow{2}{*}{Centralized Serving}
& Qwen3-32B & $97$ & $80$ & $54$ & $41$ \\
& Qwen3-14B & $198$ & $165$ & $109$ & $86$ \\
& \black{Llama-70B} & \black{$85$} & \black{$26$} & \black{$13$} & \black{$4$} \\
\midrule
\multirow{2}{*}{\acronym}
& Qwen3-32B & $164$ & $142$ & $103$ & $86$ \\
& Qwen3-14B & $245$ & $223$ & $194$ & $154$ \\
& \black{Llama-70B} & \black{$128$} & \black{$64$} & \black{$22$} & \black{$9$} \\
\midrule
\multirow{2}{*}{\textit{Speedup vs.\ SLED}}
& Qwen3-32B &
$1.98 \times$ &
$3.38\times$ &
$3.81\times$ &
$4.10\times$ \\
& Qwen3-14B &
$2.06\times$ &
$2.53\times$ &
$2.73\times$ &
$2.96\times$ \\
& \black{Llama-70B} &
\black{$2.67\times$} &
\black{$2.56\times$} &
\black{$3.14\times$} &
\black{$3.00\times$} \\
\midrule
\multirow{2}{*}{\textit{Speedup vs.\ Centralized}}
& Qwen3-32B &
$1.69\times$ &
$1.78\times$ &
$1.91\times$ &
$2.10\times$ \\
& Qwen3-14B &
$1.24\times$ &
$1.35\times$ &
$1.78\times$ &
$1.79\times$ \\
& \black{Llama-70B} &
\black{$1.51\times$} &
\black{$2.46\times$} &
\black{$1.69\times$} &
\black{$2.25\times$} \\
\bottomrule
\end{tabular}
\end{table}

\subsubsection{System goodput}
We further compare \emph{system goodput}, defined as the aggregate throughput of \emph{verified and committed} tokens across all devices (tokens/s), i.e., excluding rejected draft tokens and uncommitted intermediate outputs. Table~\ref{tab:system_goodput} reports that \acronym consistently delivers significantly higher goodput than both baselines under the same verifier budget. This improvement is driven by three mechanisms: (i) the rejection predictor reduces wasted speculative work, increasing the fraction of drafted tokens that become committed outputs; (ii) prefix-cache reuse with PageAttention accelerates consecutive verifications by reusing KV states, improving verifier-side batch efficiency; and (iii) offloading token generation to edge devices frees the GPU verifier from autoregressive decoding. Together, these effects increase the amount of useful output produced per unit verifier time, especially near saturation where baseline goodput collapses due to queueing and interference.

\begin{table}[h]
\centering
\scriptsize
\setlength{\tabcolsep}{6pt}
\renewcommand{\arraystretch}{1.15}
\caption{System goodput comparison.}
\label{tab:system_goodput}
\begin{tabular}{lccccc}
\toprule
\textbf{Metric} &
\textbf{Target Model} &
\textbf{\acronymsled} &
\textbf{Centralized Serving} &
\textbf{\acronym} \\
\midrule
\multirow{2}{*}{Whole System Goodput}
& Qwen3-32B & 170.82 \textit{tokens/s} & 326.12 \textit{tokens/s} & 631.86 \textit{tokens/s} \\
& Qwen3-14B & $276.38$ \textit{tokens/s} & $552.34$ \textit{tokens/s} & $1044.92$ \textit{tokens/s} \\
& \black{Llama-70B} & \black{$103.25$ \textit{tokens/s}} & \black{$191.52$ \textit{tokens/s}} & \black{$348.40$ \textit{tokens/s}} \\
\bottomrule
\end{tabular}
\end{table}

\subsection{SLO Analysis}
We next examine how well \acronym sustains token-speed SLOs under increasing multi-tenant load. We consider four SLO classes with requirements $c\in\{2,4,6,8\}$~tokens/s (Class~1--4). In each run, we sweep the number of simultaneously served edge devices $N$ and generate multiple responses per device. A request is counted as an SLO violation if its achieved end-to-end token speed (including queueing and verification) falls below device's class requirement.

\begin{figure*}[t]
  \centering
  \begin{subfigure}[t]{0.25\textwidth}
    \centering
    \includegraphics[width=\linewidth]{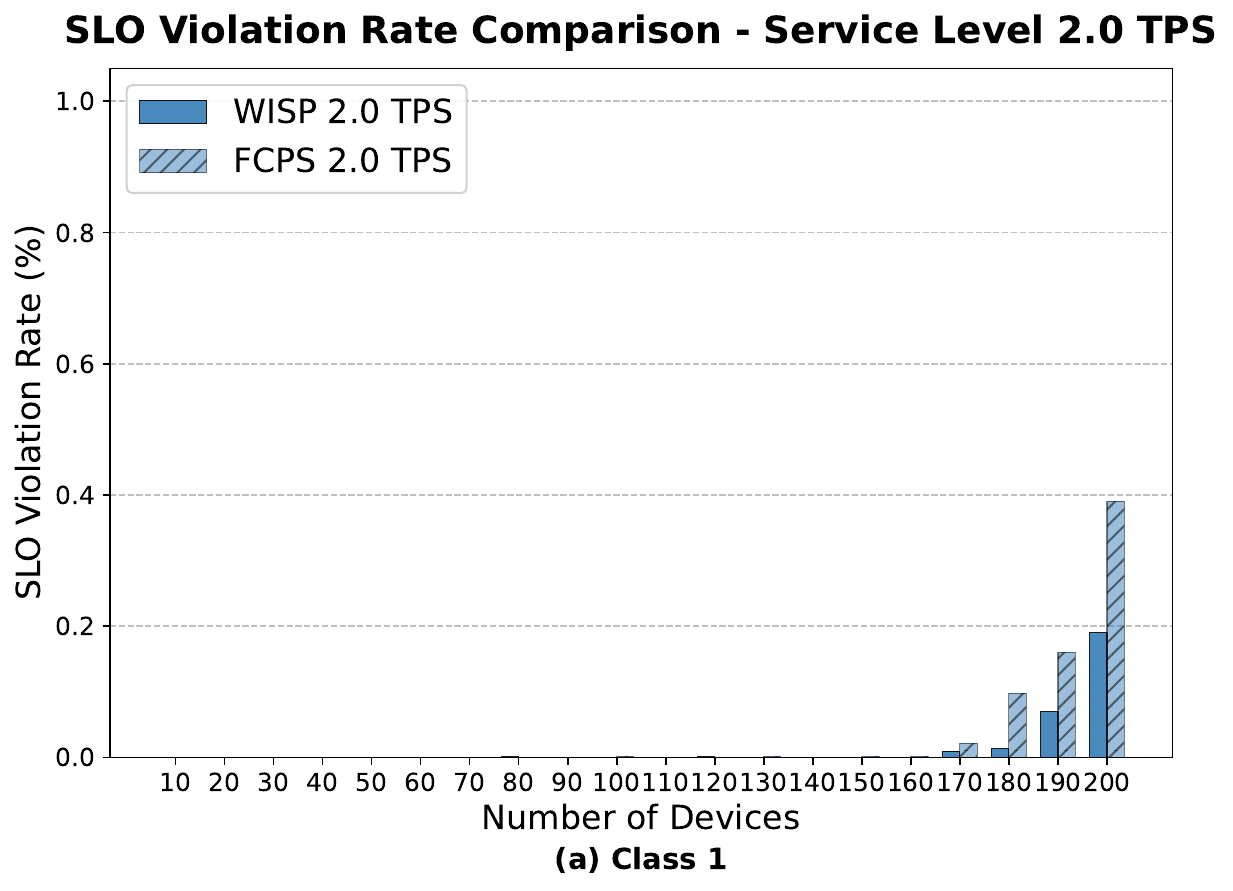}
    \caption{Class 1}
  \end{subfigure}\hfill
  \begin{subfigure}[t]{0.25\textwidth}
    \centering
    \includegraphics[width=\linewidth]{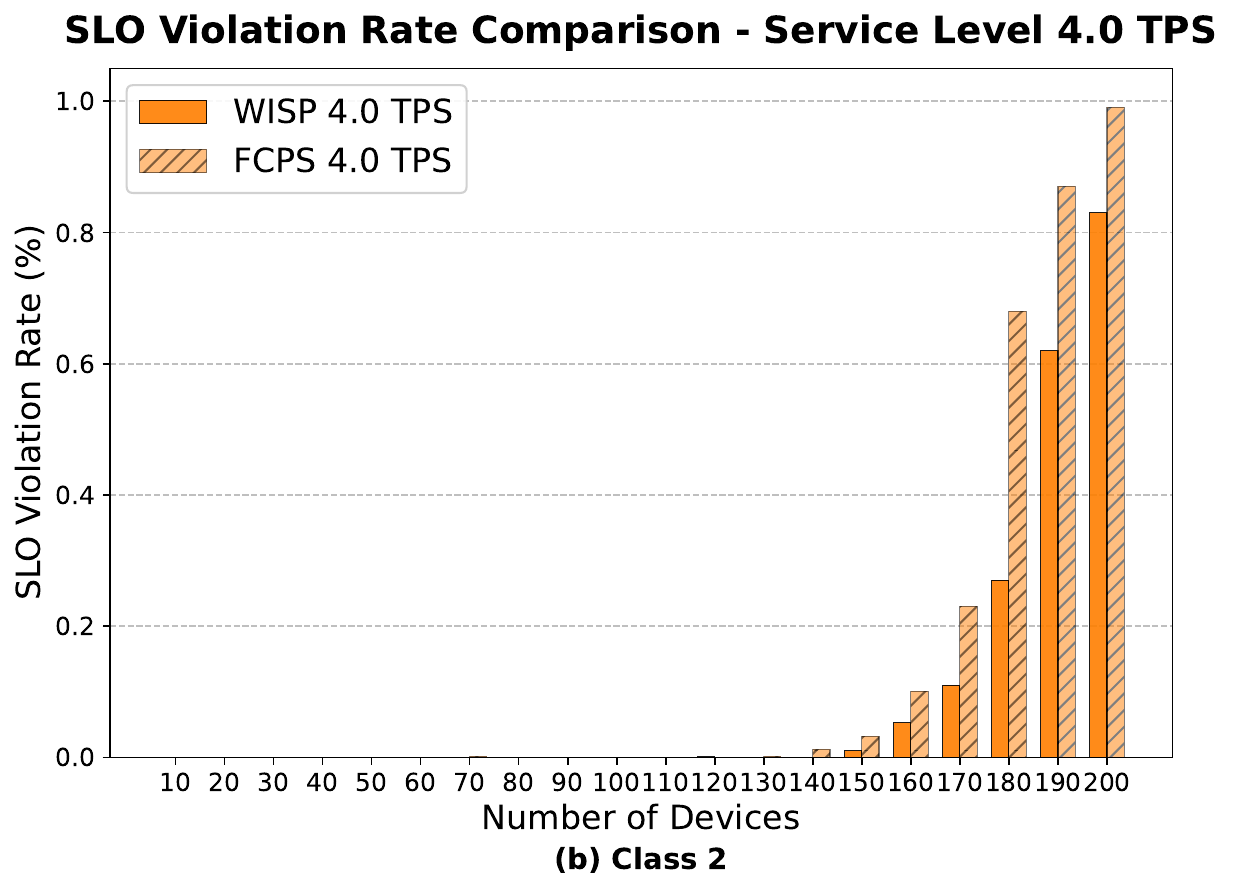}
    \caption{Class 2}
  \end{subfigure}\hfill
  \begin{subfigure}[t]{0.25\textwidth}
    \centering
    \includegraphics[width=\linewidth]{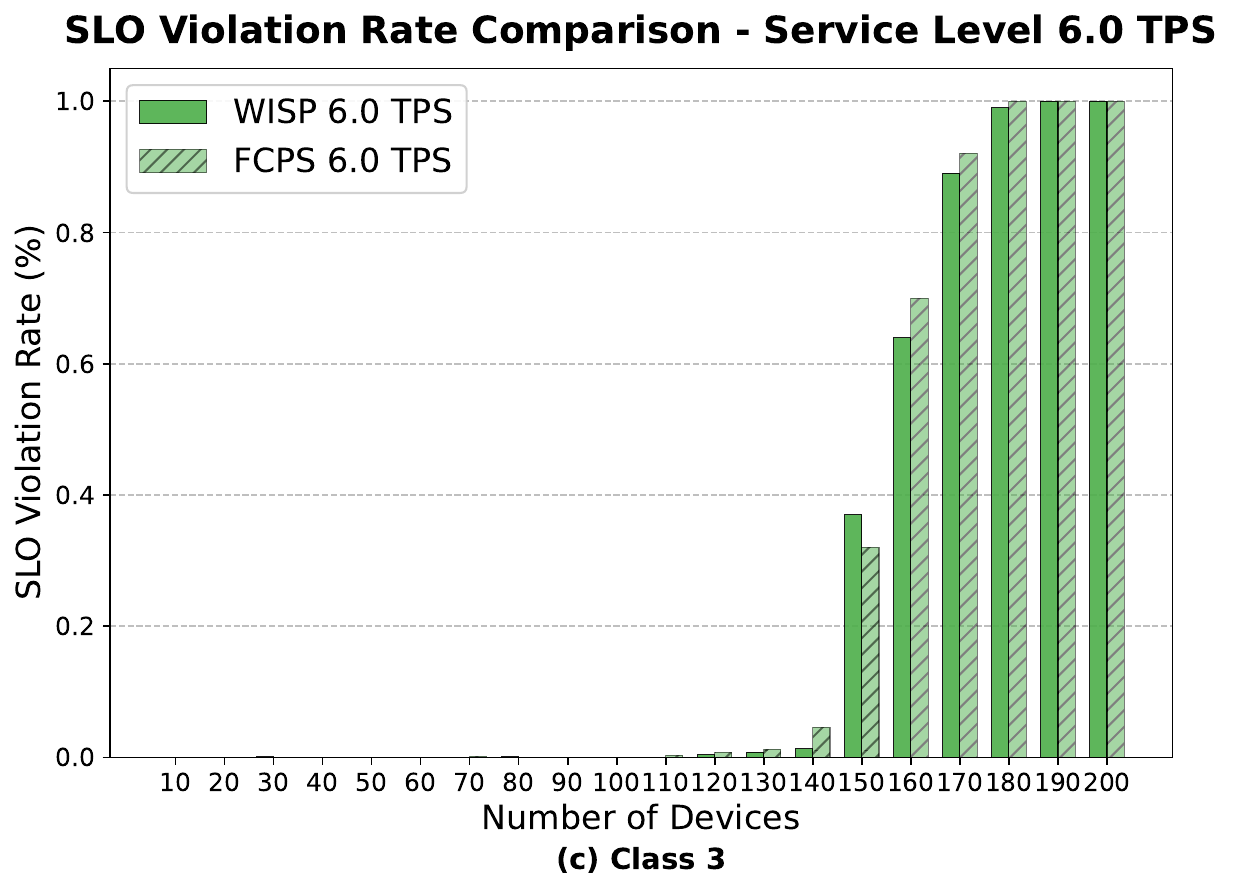}
    \caption{Class 3}
  \end{subfigure}\hfill
  \begin{subfigure}[t]{0.25\textwidth}
    \centering
    \includegraphics[width=\linewidth]{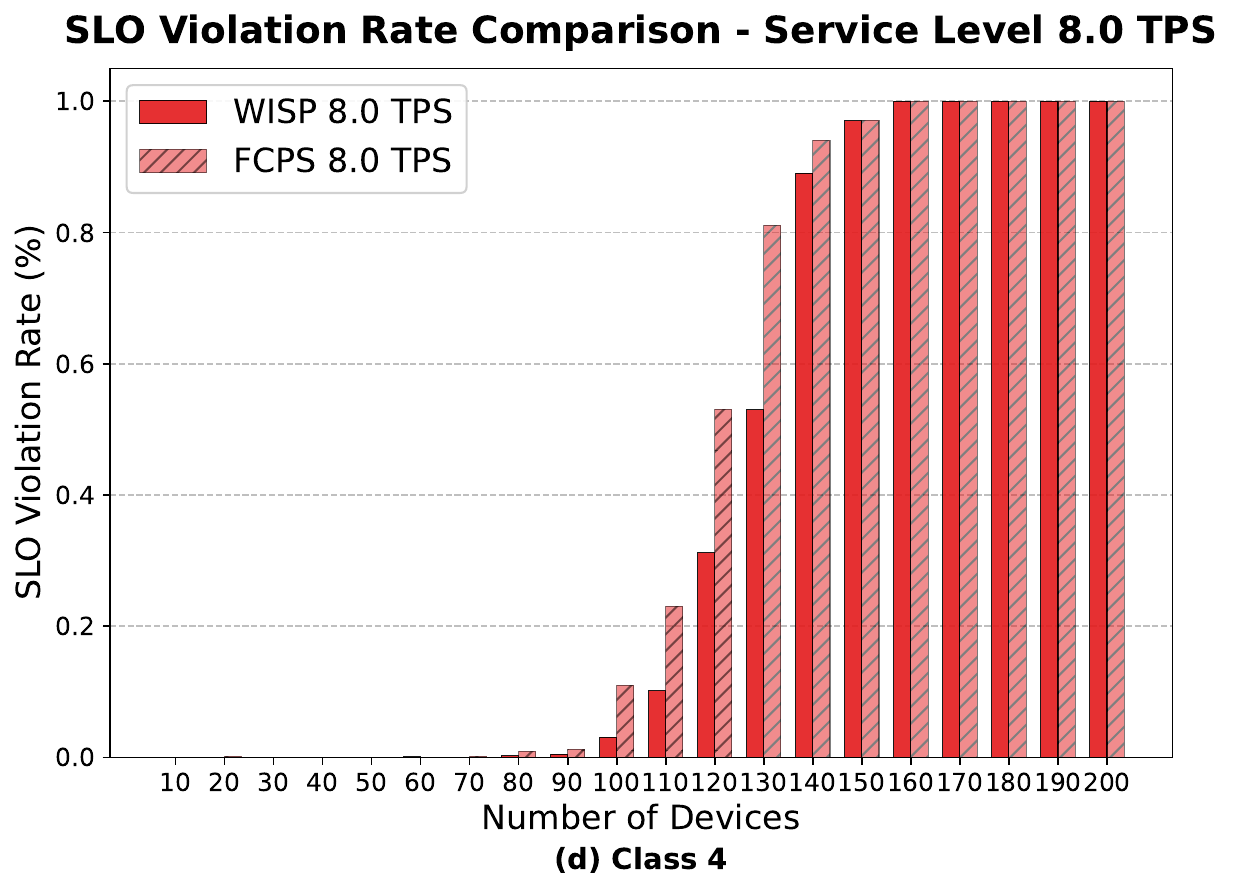}
    \caption{Class 4}
  \end{subfigure}
  \caption{Token Generation Speed and SLO Violations}
  \Description{This figure shows the SLO violation comparisons of \acronym and FCFS system under four SLO classes. The \acronym always reaches the lower SLO violations.}
  \label{fig:slo_violation_barplot:cms}
\end{figure*}

Figure.~\ref{fig:slo_violation_barplot:cms} plots the violation rate as a function of $N$ for \acronym and an FCFS verification baseline, which is implemented by disabling the scheduler in \acronym. Two patterns are clear. First, violations stay near zero at low load and then rise sharply once the verifier approaches saturation, indicating a ``knee'' where queueing and batch completion time begin to dominate. Second, \acronym consistently shifts this knee to larger $N$ and reduces the violation rate at the same load across all classes, with the largest relative gains in tighter classes (6/8~tokens/s) where latency headroom is smallest. This is because FCFS readily mixes heterogeneous verification lengths inside the same batch, amplifying head-of-line blocking and cascading delays to short requests, whereas \acronym constructs latency-feasible batches and prioritizes urgency using the online verification-time estimator, thereby containing \emph{Verification Interference} and preventing runaway queue growth.

\begin{figure*}[t!]
    \centering
    \includegraphics[width=0.5\textwidth]{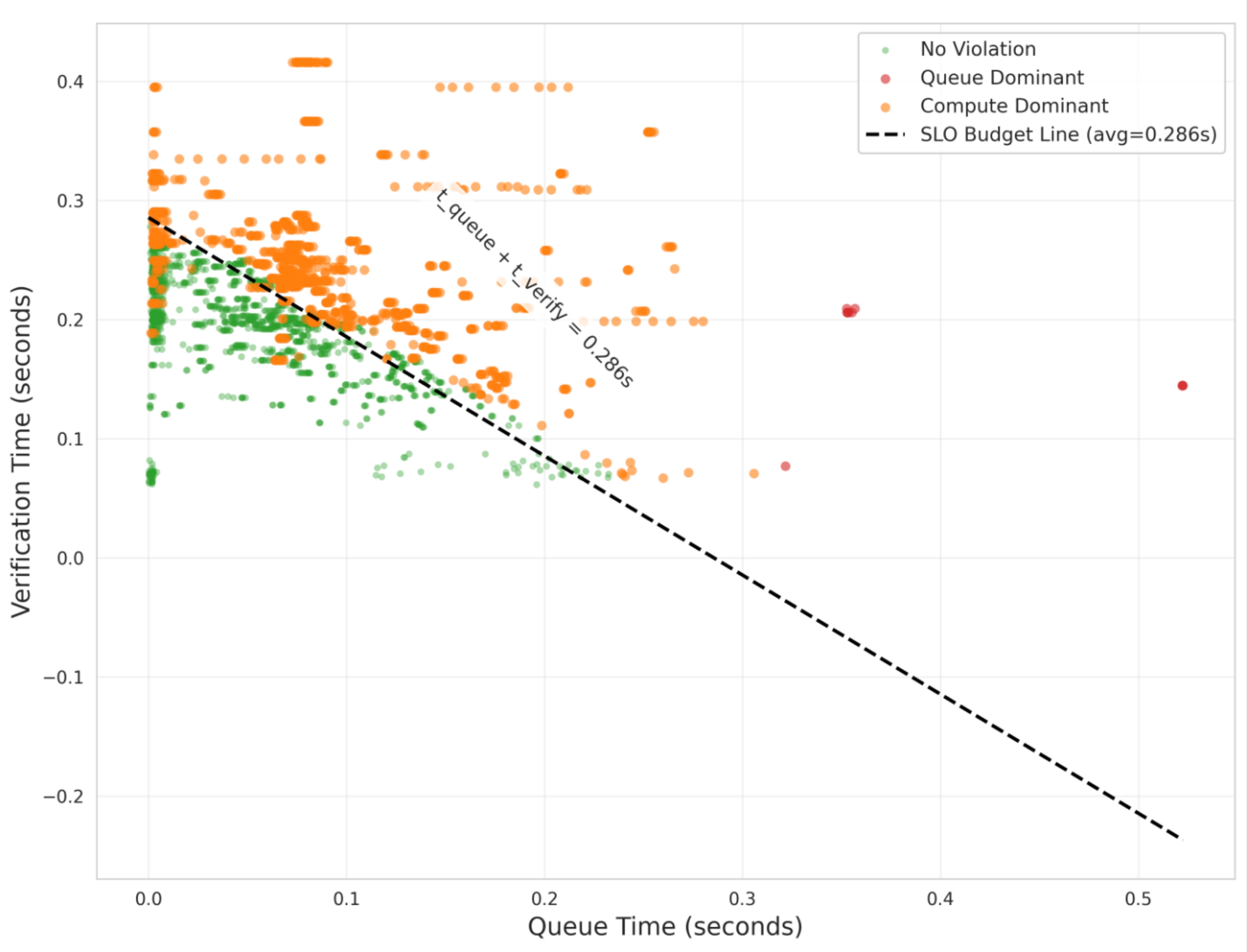}
    \caption{Violation attribution in the queue--compute plane under \acronym.}
    \Description{Scatter plot of per-verification events with queueing delay on the x-axis and verification execution time on the y-axis. Points are colored by outcome (no violation) and by spike-based attribution among violated events (queue-dominant vs compute-dominant), with a dashed reference budget line.}
    \label{fig:violation_attribution_scatter}
\end{figure*}

\paragraph{Violation attribution in the queue--compute plane.}
We analyze \acronym's verification-SLO violations using per-verification logs from the concurrent device simulator.
Each verification event records the queueing delay $t_{\text{queue}}$ (time spent waiting to be scheduled into a batch) and the measured GPU verification time $t_{\text{verify}}$ (batch execution time attributed to the request).
We plot all events in the $(t_{\text{queue}}, t_{\text{verify}})$ plane (Figure.~\ref{fig:violation_attribution_scatter}) and color points by whether they violate the SLO.

Among violated events, we label an event as \emph{compute-dominant} if its verification time exhibits a clear spike relative to recent server behavior.
Let $\mathrm{MA}_W(t_{\text{verify}})$ be the sliding mean over the most recent $W$ batches (or events); we define
\begin{equation}
\rho \triangleq \frac{t_{\text{verify}}}{\mathrm{MA}_W(t_{\text{verify}})} ,
\end{equation}
and classify a violated event as compute-dominant if $\rho > \tau$ (e.g., $\tau{=}1.5$). Otherwise, violated events are labeled \emph{queue-dominant}, indicating they occur under a near-baseline $t_{\text{verify}}$ and are more consistent with backlog-induced waiting.
The figure highlights that most violations coincide with elevated $t_{\text{verify}}$ spikes even when $t_{\text{queue}}$ is moderate, while a smaller fraction arises from large $t_{\text{queue}}$ without a corresponding compute spike, revealing distinct violation regimes under batched verification.

\begin{figure*}[t!]
    \centering
    \includegraphics[width=0.8\textwidth]{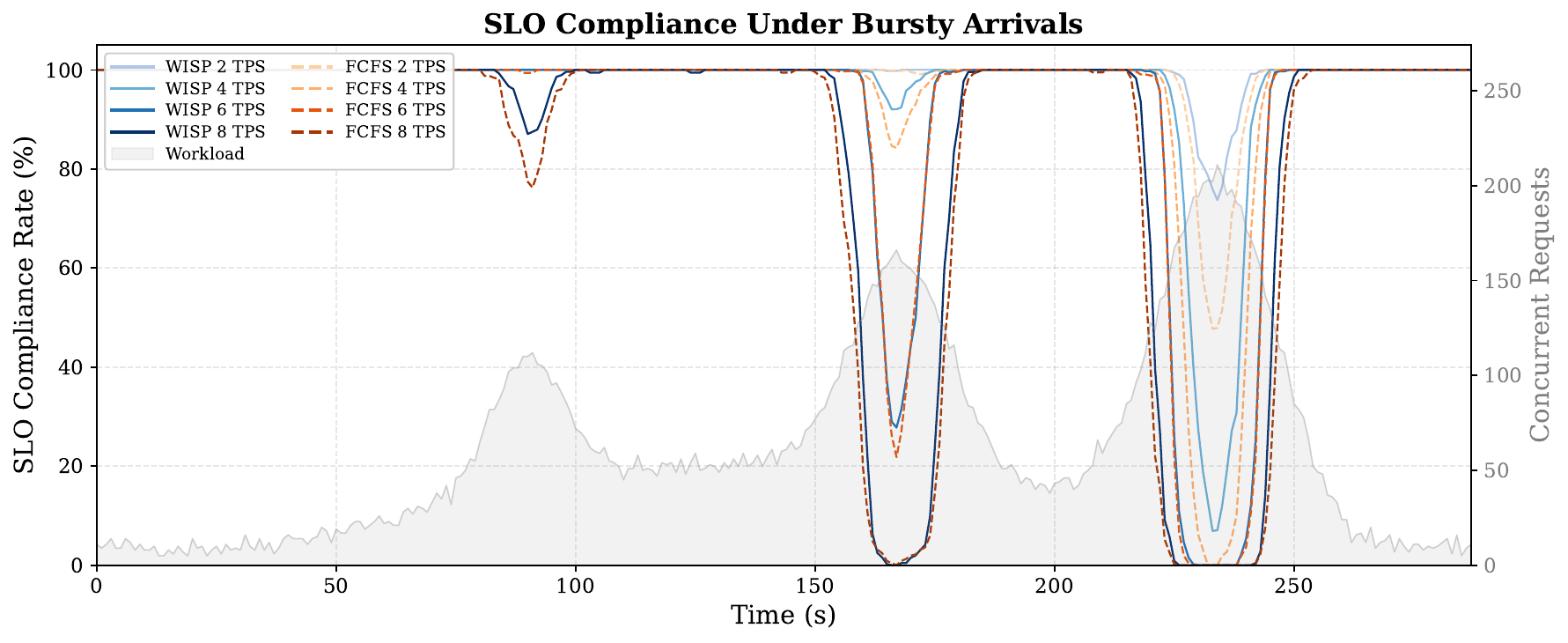}
    \caption{SLO compliance under bursty workload.}
    \Description{Plot showing SLO compliance of \acronym and FCFS system under bursty workload, for 4 service levels}
    \label{fig:bursty}
\end{figure*}

\textcolor{black}{\paragraph{Stress-test under bursty arrivals.}
To evaluate robustness under bursty traffic conditions, we construct a synthetic workload trace consisting of three burst events, with peak loads of approximately 110, 155, and 215 concurrent requests, superimposed on a diurnal baseline ranging from 10 to 55 requests. Under this workload, we compare \acronym against the FCFS baseline across four SLO classes, using Qwen3-32B as the target model. Figure~\ref{fig:bursty} reports the instantaneous SLO compliance rate for all four service classes. During off-peak periods, both systems maintain near-perfect compliance. The performance gap becomes evident during burst events. At the first burst, with a peak of 110 concurrent requests, both systems remain robust for the 2--6 TPS SLO classes, while for the most stringent 8 TPS class, \acronym improves the SLO compliance rate by 18 percentage points over FCFS. As the burst intensity increases, compliance degrades for both systems; however, \acronym consistently demonstrates stronger resilience. At the largest burst, with 215 concurrent requests, \acronym still outperforms FCFS by 17 and 6 percentage points for the 2 TPS and 4 TPS service classes, respectively. Across all service classes, \acronym also recovers to full compliance more rapidly after each burst subsides, demonstrating that its priority-aware scheduling strategy provides substantial robustness under bursty arrivals, particularly for latency-sensitive services.}

\subsection{Intelligent Drafting Controller Evaluations}
This subsection evaluates the intelligent drafting controller---in particular, the \emph{stop-at-first-predicted-rejection} policy---to verify that it reduces wasted drafting while preserving speculative coverage. We (i) benchmark candidate rejection predictors under an operating point that penalizes false acceptance of truly rejected tokens (which directly causes over-drafting), and (ii) run ablations that toggle the predictor ON/OFF to measure downstream effects on draft-token acceptance and end-to-end system goodput. Overall, the MLP offers the best trade-off for our objective by cutting FPR to 0.425 (about $2\times$ lower than tree ensembles), and enabling the predictor increases draft acceptance by $34\%$--$54\%$ and system goodput by $20\%$--$30\%$.

\begin{table}[t]
\centering
\scriptsize
\renewcommand{\arraystretch}{1.15}
\setlength{\tabcolsep}{6pt}
\begin{tabular}{lcccccc}
\toprule
Predictor & Acc. & AUC & Rec$_1$ & Spec. & FPR & BalAcc \\
\midrule
MLP (3-layer) & 0.7510 & 0.7727 & 0.8011 & 0.5750 & 0.4250 & 0.6881 \\
LightGBM      & 0.7431 & 0.7939 & 0.9396 & 0.1404 & 0.8596 & 0.5400 \\
Random Forest & 0.7459 & 0.7964 & 0.9396 & 0.1517 & 0.8483 & 0.5457 \\
XGBoost       & 0.7528 & 0.7978 & 0.9322 & 0.2022 & 0.7978 & 0.5672 \\
\bottomrule
\end{tabular}
\caption{\textbf{Predictor operating point for stop-at-first-predicted-rejection.}
Rec$_1$ is recall on the Accepted (1) class (speculative coverage), while Spec./FPR quantify rejection detection on the Rejected (0) class (waste control).
BalAcc $=(\text{Rec}_1+\text{Spec.})/2$.}
\label{tab:predictor:main}
\end{table}

\paragraph{Evaluation and model choice.}
Our dataset is acceptance-dominated, so overall accuracy can remain high even for predictors biased toward predicting acceptance.
However, under our \emph{stop-at-first-predicted-rejection} drafting policy, \emph{false positives on truly rejected tokens} directly cause over-drafting and WDT, because they allow the device to draft past the first true rejection.
Therefore, we prioritize rejection detection (Spec./FPR) over majority-class recall.
As shown in Table~\ref{tab:predictor:main}, tree-based models achieve high Accepted recall (Rec$_1\!\approx\!0.93$--$0.94$) but exhibit very poor specificity (Spec.$=0.14$--$0.20$, i.e., FPR $=0.80$--$0.86$), which would translate into frequent over-drafting when an early rejection occurs.
In contrast, the MLP reduces FPR to 0.425 (a $\sim$2$\times$ reduction), yielding the best balanced accuracy and better aligning with our objective of minimizing WDT; we therefore adopt the MLP in \acronym.
Detailed per-class counts and normalized confusion matrices are deferred to Appendix~\ref{app:predictor_detail}.

\begin{figure*}[t!]
    \centering
    \includegraphics[scale=0.45]{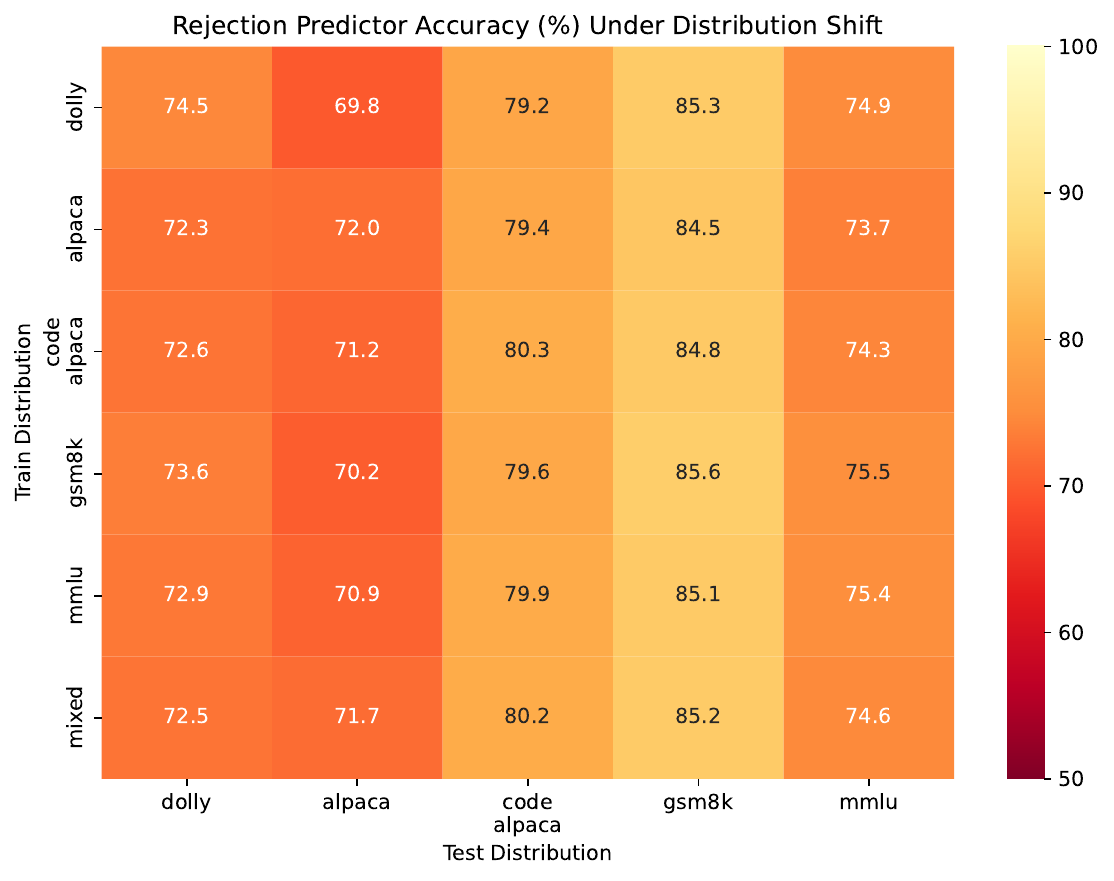}
    \caption{Rejection predictor distribution shift}
    \Description{Accuracies of rejection predictor under distribution shift}
    \label{fig:distribution-shift}
\end{figure*}

\textcolor{black}{\paragraph{Robustness under distribution shift.} To evaluate the sensitivity of the rejection predictor to the prompt distribution used during training, we collected speculative decoding trajectories from five diverse prompt datasets: Dolly\cite{conover2023dolly} (general instruction), Alpaca\cite{taori2023alpaca} (instruction-following), CodeAlpaca\cite{chaudhary2023codealpaca} (code generation), GSM8K\cite{cobbe2021gsm8k} (mathematical reasoning), and MMLU\cite{hendrycks2021mmlu} (academic QA), with Qwen3-1.7B as the draft model and Qwen3-32B as the target model. For each dataset, we trained an MLP classifier (hidden layers: 512-128-64), and evaluated it on all five test distributions. As shown in Fig.~\ref{fig:distribution-shift}, the predictor exhibits strong robustness to distribution shift: for any given test distribution, the accuracy varies by at most 2.2 percentage points regardless of the training source. This uniformity across rows indicates that the input features, including draft confidence, entropy, margin, and standard deviation, capture distribution-agnostic statistical properties of token acceptance rather than content-specific patterns. Notably, a model trained on pooled data from all five distributions performs within 1.9 percentage points of the best in-distribution model on every test set, suggesting that online adaptation through diverse data aggregation is a practical and effective strategy that does not require retraining on the target distribution.}

\subsubsection{Predictor ablations}
We isolate the contribution of the rejection predictor by turning it \emph{ON/OFF} while keeping all other components unchanged. We report two effects: (i) how the predictor changes the \emph{draft token acceptance rate} (a proxy for wasted verification work), and (ii) how this translates into \emph{end-to-end goodput} under multi-device load.

\paragraph{Draft acceptance rate (w/ vs. w/o predictor).}
Table~\ref{tab:acceptance_ablation} compares the fraction of drafted tokens that are ultimately accepted by the verifier for four draft models. Enabling the predictor increases acceptance from $0.31\!\to\!0.48$, $0.42\!\to\!0.63$, $0.47\!\to\!0.66$, and $0.53\!\to\!0.71$ for Qwen3-\{0.6B, 1.7B, 4B, 8B\}, yielding $34\%$--$54\%$ relative improvements. The gains are larger for smaller draft models because they exhibit lower baseline alignment with the target model.

\begin{table}[h]
\centering
\scriptsize
\setlength{\tabcolsep}{6pt}
\renewcommand{\arraystretch}{1.15}
\caption{Ablation of draft token acceptance rate with and without the predictor.}
\label{tab:acceptance_ablation}
\begin{tabular}{lcccc}
\toprule
\textbf{Draft Models} &
\textbf{Qwen3-0.6B} &
\textbf{Qwen3-1.7B} &
\textbf{Qwen3-4B} &
\textbf{Qwen3-8B} \\
\midrule
Predictor: OFF & 0.31 & 0.42 & 0.47 & 0.53 \\
Predictor: ON  & 0.48 & 0.63 & 0.66 & 0.71 \\
Improvement & 54.0\% $\uparrow$ & 50.0\% $\uparrow$& 44.7\% $\uparrow$& 34.0\% $\uparrow$\\
\bottomrule
\end{tabular}
\end{table}

\paragraph{System goodput (w/ vs. w/o predictor).}
Table~\ref{tab:goodput_ablation} reports aggregate system goodput (verified committed tokens/s) when serving $N\in\{2,4,8,16\}$ concurrent devices. With the predictor enabled, goodput improves by $20.5\%$--$30.0\%$, and the relative gain increases with $N$. This scaling trend indicates that the predictor primarily helps by \emph{reducing verifier-side load}: fewer rejected draft tokens means fewer wasted verification FLOPs and less batch elongation, which directly mitigates queue build-up and head-of-line blocking under contention. Consequently, near saturation (e.g., $N{=}16$), the saved verification work converts more efficiently into committed throughput.

\begin{table}[h]
\centering
\scriptsize
\setlength{\tabcolsep}{6pt}
\renewcommand{\arraystretch}{1.15}
\caption{Ablation of system goodput with and without the predictor.}
\label{tab:goodput_ablation}
\begin{tabular}{lcccc}
\toprule
\textbf{Number of devices} & \textbf{2} & \textbf{4} & \textbf{8} & \textbf{16} \\
\midrule
Predictor: OFF & 14.23 & 21.12 & 56.12 & 102.13 \\
Predictor: ON  & 17.14 & 26.43 & 70.43 & 132.80 \\
Improvement (\%) & 20.45 & 25.14 & 25.49 & 30.03 \\
\bottomrule
\end{tabular}
\end{table}

\subsection{Accuracy and Calibration of the Verification-Time Estimator}
\label{sec:eval_time_estimator}
We now evaluate the fidelity of our additive verification-time estimator, which serves as the runtime primitive for (i) predicting batch verification latency during incremental batch construction and (ii) enforcing SLO-feasibility with minimal conservatism.

Table~\ref{tab:latency_metrics} summarizes the model performance. On the training set, the model achieves $R^2 = 0.986$ with MAE of 3.85 ms and Mean Absolute Percentage Error (MAPE) of 8.91\%. On the held-out test set, performance improves slightly to $R^2 = 0.992$ with MAE of 3.38 ms and MAPE of 4.93\%, demonstrating excellent generalization. The $R^2$ difference of only 0.006 between training and test sets confirms that the model does not overfit. Cross-validation on the training set yields mean $R^2 = 0.982 \pm 0.008$, indicating stable parameter estimates across different data subsets.

\begin{table}[t]
\scriptsize
\centering
\caption{Latency model performance metrics}
\label{tab:latency_metrics}
\begin{tabular}{lrr}
\toprule
Metric & Training Set & Test Set \\
\midrule
$R^2$ & 0.986 & 0.992 \\
Adjusted $R^2$ & 0.985 & 0.991 \\
RMSE (ms) & 5.35 & 4.49 \\
MAE (ms) & 3.85 & 3.38 \\
MAPE (\%) & 8.91 & 4.93 \\
Max Error (ms) & 22.31 & 14.59 \\
\bottomrule
\end{tabular}
\end{table}

\begin{figure*}[t!]
  \centering
  \begin{subfigure}[t]{0.33\textwidth}
    \centering
    \includegraphics[width=\linewidth]{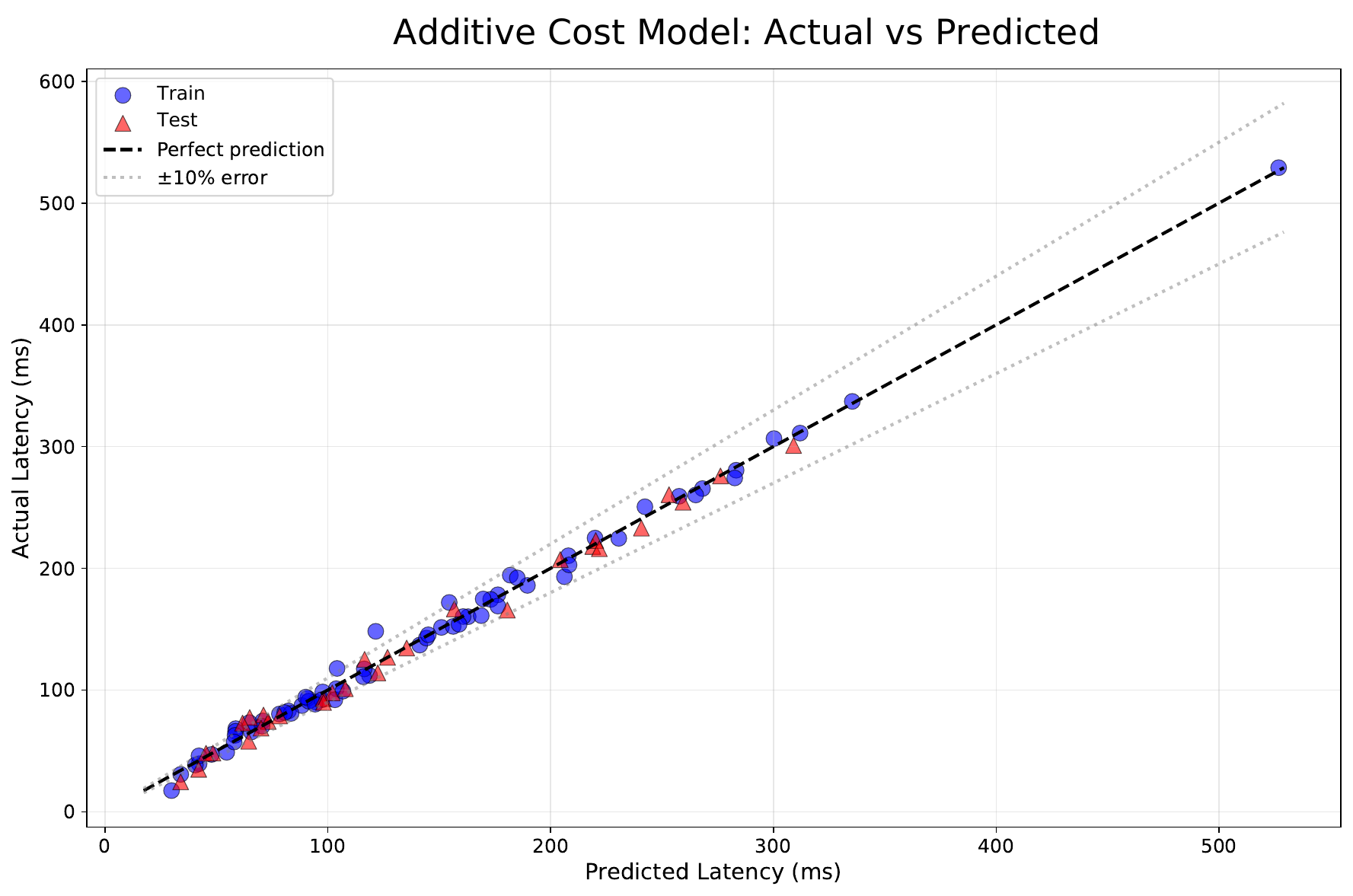}
    \caption{Actual vs. predicted}
    \label{fig:additive_actual_vs_predicted}
  \end{subfigure}\hfill
  \begin{subfigure}[t]{0.67\textwidth}
    \centering
    \includegraphics[width=\linewidth]{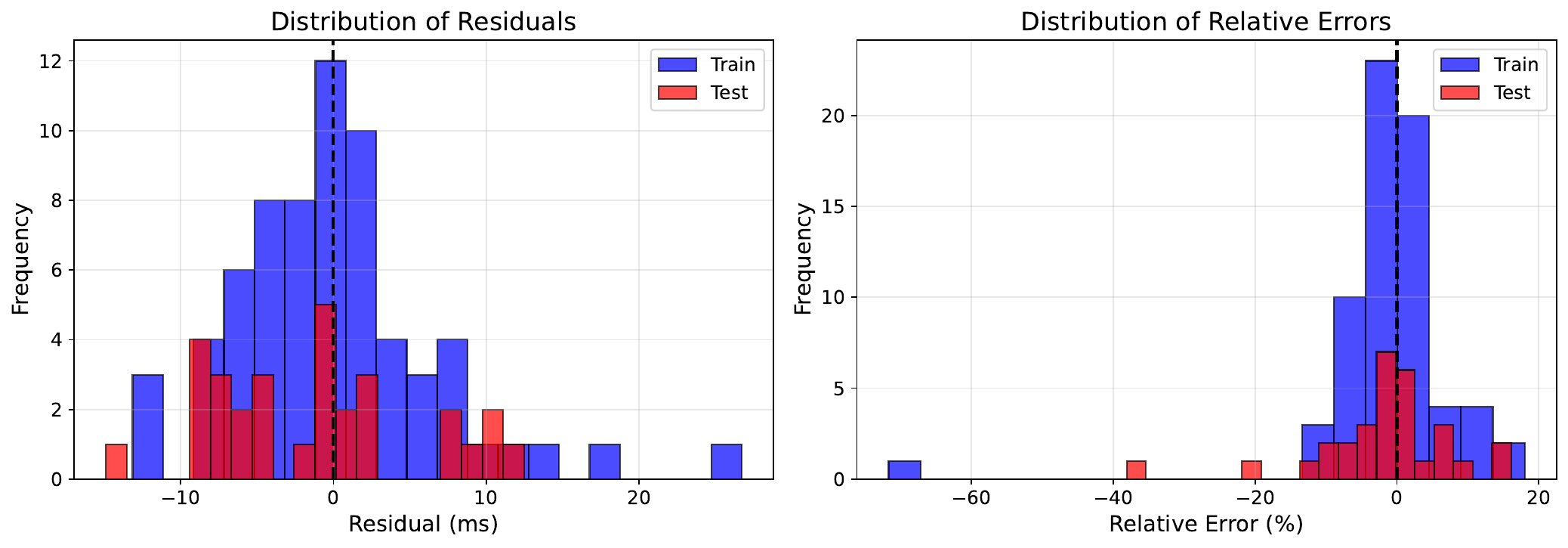}
    \caption{Residual diagnostics}
    \label{fig:additive_error_distribution}
  \end{subfigure}\hfill

  \caption{Additive verification-latency predictor: end-to-end fit and residual diagnostics.}
  \Description{Left: actual vs.\ predicted batch verification latency. Middle: residual and relative-error distributions. Right: <brief description of the third figure>.}
  \label{fig:additive_predictor_three_across}
\end{figure*}

Figure.~\ref{fig:additive_actual_vs_predicted} compares the predicted batch latency against the measured latency. The scatter closely follows the $y=x$ reference line across a wide latency range, with the majority of samples lying within the $\pm 10\%$ envelope, indicating that the additive form captures the dominant scaling trends of verification cost. This near-identity agreement supports using the estimator online to anticipate batch completion time and to guide SLO-aware admission and batching decisions.

Beyond average accuracy, a scheduler requires the estimator to be well-calibrated---i.e., free of systematic bias that would consistently under-predict latency and cause deadline misses. Figure.~\ref{fig:additive_error_distribution} therefore reports the residual and relative-error distributions. Residuals and relative errors are tightly centered around zero for both train and test splits, implying no consistent under- or over-estimation across the evaluated settings. Taken together, these diagnostics indicate that the estimator is reliable for mean behavior, while motivating the use of a small guard time (or slack factor) when enforcing hard deadlines so that rare tail deviations do not translate into SLO violations in production.

\section{Conclusion and Future Work} \label{sec:conclusion}

\acronym rethinks distributed speculative decoding as an end-to-end serving system and, for the first time, pinpoints two scaling limits: WDT on edge devices and \emph{Verification Interference} inside GPU micro-batches. It mitigates both via co-designed control and scheduling: a lightweight rejection predictor stops drafting near the first likely rejection to curb over-drafting, while the server enforces token-speed SLO time budgets with interference-aware batching (an urgent fast path plus utility-based filling) under online feasibility checks enabled by a simple verification-time estimator that captures linear, attention, and cache effects. Combined with PageAttention and prefix-cache KV reuse across consecutive verifications, \acronym increases system capacity and goodput under heterogeneous workloads.

\begin{acks}
This material is based on work supported by the National Science Foundation under Grants No. 2315851 and Major Grants Initiative Program, Virginia Tech College of Engineering.
\end{acks}

\bibliographystyle{ACM-Reference-Format}
\bibliography{sample-base}

\appendix

\section{Proof of Theorem 1}
\begin{proof}
Consider one speculate--verify iteration with a maximum speculative window $K_{\max}$.
The draft model proposes tokens $(y_1,\dots,y_{K_{\max}})$, and the verifier (target model) checks them individually.
Let
\begin{equation}
A_i \in \{0,1\},\qquad A_i=1~\text{iff token }y_i\text{ is accepted by the target model.}
\end{equation}
Define the first rejection index $R$ and the acceptance length $L$ as
\begin{equation}
R \triangleq \min\{i\in\{1,\dots,K_{\max}\}: A_i=0\},\qquad
L \triangleq R-1,
\end{equation}
with the convention $R=K_{\max}+1$ (hence $L=K_{\max}$) if no rejection occurs.

Now introduce a binary predictor $g_\theta$ that, for each step $i$, outputs
\begin{equation}
\hat Z_i \triangleq g_\theta(\text{features}_i)\in\{0,1\},
\end{equation}
where $\hat Z_i=1$ means ``predict accepted'' and $\hat Z_i=0$ means ``predict rejected''.
Under the \emph{stop-at-first-predicted-rejection} launch rule, the device drafts until the first predicted reject:
\begin{equation}
K_\theta \triangleq \max\bigl\{k\in\{0,\dots,K_{\max}\}:\hat Z_1=\cdots=\hat Z_k=1\bigr\}.
\end{equation}
The number of wasted draft tokens (over-drafted tokens beyond the true accepted prefix) is
\begin{equation}
W_\theta \triangleq (K_\theta - L)^+ = \max\{0, K_\theta - L\}.
\end{equation}

\paragraph{Step 1: Waste can only occur after the first true rejection.}
If $R=K_{\max}+1$, then $L=K_{\max}$ and, since $K_\theta\le K_{\max}$ by construction, we have $W_\theta=(K_\theta-L)^+=0$.
Hence, it suffices to consider the event $\{R=r\}$ for some $r\in\{1,\dots,K_{\max}\}$.
On $\{R=r\}$, we have $A_r=0$ and $A_1=\cdots=A_{r-1}=1$, so $L=r-1$.
Moreover, if the predictor outputs $\hat Z_r=0$ at the first true rejection, then by the stop rule the device cannot draft beyond $r-1$ tokens, i.e.,
$K_\theta \le r-1 = L$, which implies $W_\theta=0$.
Therefore, a necessary condition for having $W_\theta>0$ is
\begin{equation}
\hat Z_R = 1 \quad \text{and} \quad A_R = 0,
\end{equation}
namely, the predictor makes a false positive (false alarm) at the first rejection index.

\paragraph{Step 2: An upper bound on the conditional expected waste.}
Fix any $r\in\{1,\dots,K_{\max}\}$ and condition on $\{R=r\}$.
Since $L=r-1$ on this event and $K_\theta\le K_{\max}$ always holds, we obtain the deterministic bound
\begin{equation}
W_\theta=(K_\theta-L)^+ \le (K_{\max}-(r-1))\,\mathbbm{1}\{\hat Z_r=1\}
= (K_{\max}-r+1)\,\mathbbm{1}\{\hat Z_r=1\}.
\end{equation}
Taking conditional expectation given $R=r$ yields
\begin{equation}
\mathbb{E}[W_\theta \mid R=r]
~\le~ (K_{\max}-r+1)\,\Pr(\hat Z_r=1 \mid R=r).
\end{equation}
Define the (position-dependent) false-positive probability at the first rejection as
\begin{equation}
\mathrm{FP}_\theta(r) \triangleq \Pr(\hat Z_r=1 \mid R=r).
\end{equation}

\paragraph{Step 3: Unconditioning and monotonicity.}
Unconditioning over $R$ and using $W_\theta=0$ when $R=K_{\max}+1$ gives
\begin{equation}
\mathbb{E}[W_\theta]
~=~ \sum_{r=1}^{K_{\max}} \Pr(R=r)\,\mathbb{E}[W_\theta \mid R=r]
~\le~ \sum_{r=1}^{K_{\max}} (K_{\max}-r+1)\,\Pr(R=r)\,\mathrm{FP}_\theta(r).
\label{eq:wdt-fp-bound}
\end{equation}

Consider two predictors $\theta$ and $\theta'$ evaluated under the same launch rule and the same underlying data-generating process for $(A_1,\ldots,A_{K_{\max}})$ (hence the same distribution of $R$).
Assume that $\theta'$ has a lower false alarm rate than $\theta$ at the first rejection, in the sense that
\begin{equation}
\mathrm{FP}_{\theta'}(r) \le \mathrm{FP}_{\theta}(r),\qquad \forall r\in\{1,\dots,K_{\max}\}.
\end{equation}
Applying \eqref{eq:wdt-fp-bound} to both predictors and comparing term by term yields
\begin{equation}
\mathbb{E}[W_{\theta'}] \le \mathbb{E}[W_{\theta}],
\end{equation}
which proves the theorem statement.

Finally, if the per-token drafting time is $\tau_d$, then the WDT satisfies
$T_{\mathrm{wdt},\theta}=\tau_d\,W_\theta$ and thus
\begin{equation}
\mathbb{E}[T_{\mathrm{wdt},\theta'}] = \tau_d\,\mathbb{E}[W_{\theta'}]
~\le~ \tau_d\,\mathbb{E}[W_{\theta}]
= \mathbb{E}[T_{\mathrm{wdt},\theta}],
\end{equation}
so the same monotonicity holds for expected WDT.
\end{proof}

\section{Performance Analysis of Rejection Predictor}
\paragraph{Training dataset generation.}
We generate a token-level supervised dataset by logging speculative decoding traces.
Each sample corresponds to a drafted token position and is labeled by the verification outcome:
a token is labeled $1$ if it is accepted and $0$ if it is the first rejected token in that speculative window.
Tokens \emph{after} the first rejection are excluded because the verifier terminates at the first mismatch, and those suffix tokens
do not affect $L$ or $W$. This matches the predictor's operational role: detecting the first imminent rejection.
The resulting dataset contains $17{,}510$ samples with $4$ features, with a class imbalance toward the positive (accepted) class.
Table~\ref{tab:dataset-stats} reports the summary statistics.

\begin{table}[h]
  \centering
  \small
  \caption{Training dataset statistics (token-level samples).}
  \label{tab:dataset-stats}
  \begin{tabular}{@{}lrr@{}}
    \toprule
    \textbf{Item} & \textbf{Count} & \textbf{Fraction} \\
    \midrule
    Total samples & 17{,}510 & 100.00\% \\
    Positive (accepted) & 12{,}820 & 73.22\% \\
    Negative (rejected) & 4{,}690 & 26.78\% \\
    Feature dimension & 5 & -- \\
    \bottomrule
  \end{tabular}
\end{table}

\paragraph{Model candidates and selection rationale.}
We evaluate four lightweight classifier families that can run efficiently on edge CPUs:
(i) gradient-boosted trees (XGBoost), (ii) histogram-based gradient boosting (LightGBM),
(iii) random forests, and (iv) a compact multi-layer perceptron (MLP).
All models consume the same 5-D feature vector to ensure a fair accuracy--overhead comparison.

\paragraph{Training protocol and hyperparameter scanning.}
We split the dataset into training/validation partitions and use the validation split to select hyperparameters. For tree-based models, we train with standard cross-entropy objectives and early stopping (when supported) to prevent overfitting; for the MLP, we train with binary cross-entropy loss and Adam, and employ mini-batch training with learning-rate and dropout tuning. We perform systematic hyperparameter tuning for each model family, and Table~\ref{tab:tuning-space} summarizes representative tuning dimensions for reproducibility. In particular, the MLP scan enumerates multiple hidden-layer shapes, learning rates, batch sizes, and regularization options (dropout, batch normalization), while for tree ensembles we scan estimator count, depth, learning rate, and subsampling knobs. The best configuration per family is selected by validation performance and then exported as the deployment artifact.

\begin{table}[t]
  \centering
  \small
  \caption{Representative hyperparameter search space for predictor tuning.}
  \label{tab:tuning-space}
  \begin{tabularx}{\linewidth}{@{}lX@{}}
    \toprule
    \textbf{Model} & \textbf{Scanned hyperparameters (examples)} \\
    \midrule
    XGBoost &
      \makecell[l]{\texttt{n\_estimators}, \texttt{max\_depth}, \texttt{learning\_rate}, \texttt{subsample}, \texttt{colsample\_bytree},\\
      \texttt{min\_child\_weight}, \texttt{gamma}, \texttt{reg\_lambda}, \texttt{reg\_alpha}} \\
    LightGBM &
      \makecell[l]{\texttt{num\_leaves}, \texttt{max\_depth}, \texttt{learning\_rate}, \texttt{n\_estimators}, \texttt{subsample},\\
      \texttt{colsample\_bytree}, \texttt{min\_child\_samples}, \texttt{reg\_lambda}, \texttt{reg\_alpha}} \\
    Random Forest &
      \makecell[l]{\texttt{n\_estimators}, \texttt{max\_depth}, \texttt{min\_samples\_split}, \texttt{min\_samples\_leaf}, \texttt{max\_features}} \\
    MLP &
      \makecell[l]{Hidden sizes (e.g., $[512,128,64]$), learning rate ($10^{-5}$--$10^{-3}$), batch size (e.g., 32--256),\\
      dropout (0--0.3), batch normalization (on/off), max epochs and early stopping} \\
    \bottomrule
  \end{tabularx}
\end{table}

\subsection{Additional Predictor Results}
\label{app:predictor_detail}

\begin{table*}[t]
\centering
\small
\renewcommand{\arraystretch}{1.15}
\setlength{\tabcolsep}{7pt}
\begin{tabular}{lcccccc}
\toprule
\multirow{2}{*}{Predictor} &
\multicolumn{2}{c}{True Rejected (0)} &
\multicolumn{2}{c}{True Accepted (1)} &
\multicolumn{2}{c}{Class Support} \\
\cmidrule(lr){2-3}\cmidrule(lr){4-5}\cmidrule(lr){6-7}
& Pred 0 (TN) & Pred 1 (FP) & Pred 0 (FN) & Pred 1 (TP) & \#Rejected & \#Accepted \\
\midrule
MLP (3-layer)     &  92 (57.50\%) &  68 (42.50\%) & 112 (19.89\%) & 451 (80.11\%) & 160 & 563 \\
LightGBM          &  25 (14.04\%) & 153 (85.96\%) &  33 ( 6.04\%) & 513 (93.96\%) & 178 & 546 \\
Random Forest     &  27 (15.17\%) & 151 (84.83\%) &  33 ( 6.04\%) & 513 (93.96\%) & 178 & 546 \\
XGBoost           &  36 (20.22\%) & 142 (79.78\%) &  37 ( 6.78\%) & 509 (93.22\%) & 178 & 546 \\
\bottomrule
\end{tabular}
\caption{\textbf{Row-normalized confusion matrices} for draft-token accept/reject prediction.
Percentages are normalized within each true class (Rejected or Accepted).}
\label{tab:predictor:cm_appendix}
\end{table*}

\paragraph{Detailed confusion-matrix interpretation.}
Table~\ref{tab:predictor:cm_appendix} highlights a consistent operating-point contrast.
For the Rejected (0) class, the tree ensembles misclassify most rejected tokens as Accepted (1): their false-positive rates on Rejected tokens are 79.78--85.96\%, meaning that the predictor would often \emph{fail to stop} at the true first rejection and thus extend the speculative window into wasted suffix tokens.
The MLP substantially improves separability on this minority class: it correctly flags 57.50\% of rejected tokens (TN), reducing the Rejected-class false positives to 42.50\%.
For the Accepted (1) class, the tree-based models keep false negatives low (6.04--6.78\%), implying stronger speculative coverage when rejections are rare, whereas the MLP incurs a higher Accepted-class FN rate (19.89\%).
These per-class patterns explain the \emph{waste-vs-coverage} trade-off discussed in the main text: low FPR/high specificity controls WDT, while high Rec$_1$ favors longer speculative coverage.


\begin{figure*}[t]
  \centering
  \begin{subfigure}[t]{0.24\textwidth}
    \centering
    \includegraphics[width=\linewidth]{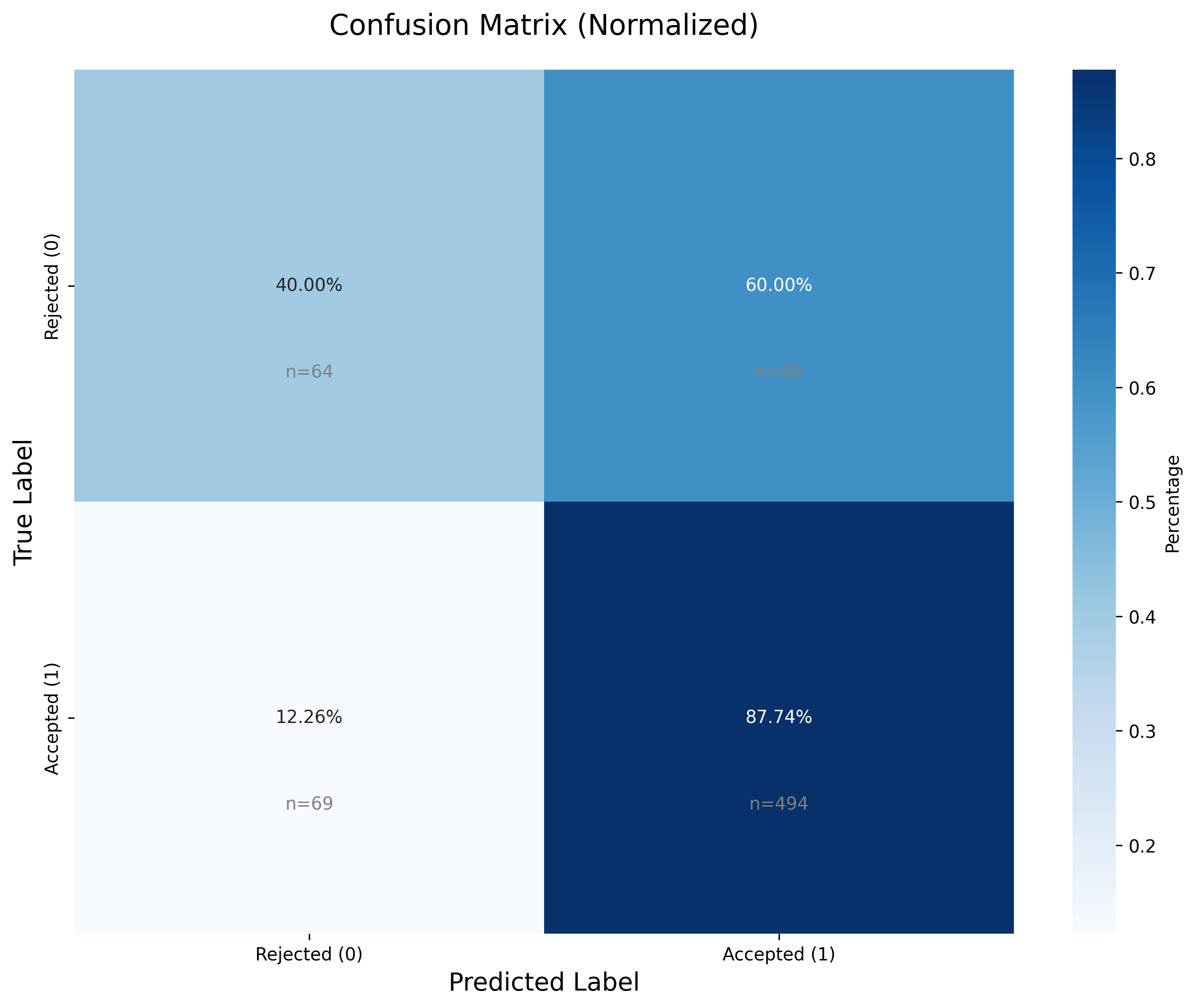}
    \caption{MLP}
  \end{subfigure}\hfill
  \begin{subfigure}[t]{0.24\textwidth}
    \centering
    \includegraphics[width=\linewidth]{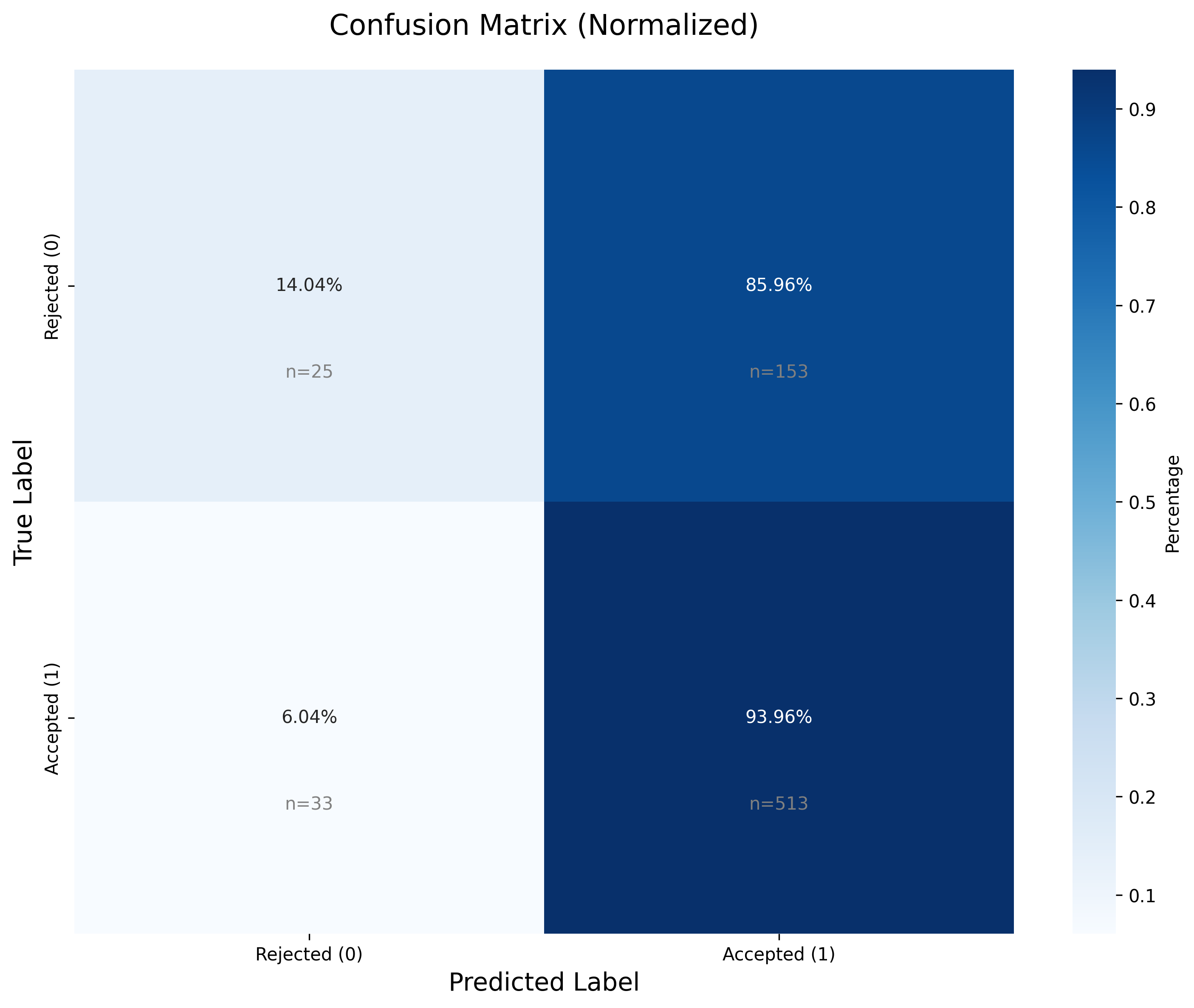}
    \caption{LightGBM}
  \end{subfigure}\hfill
  \begin{subfigure}[t]{0.24\textwidth}
    \centering
    \includegraphics[width=\linewidth]{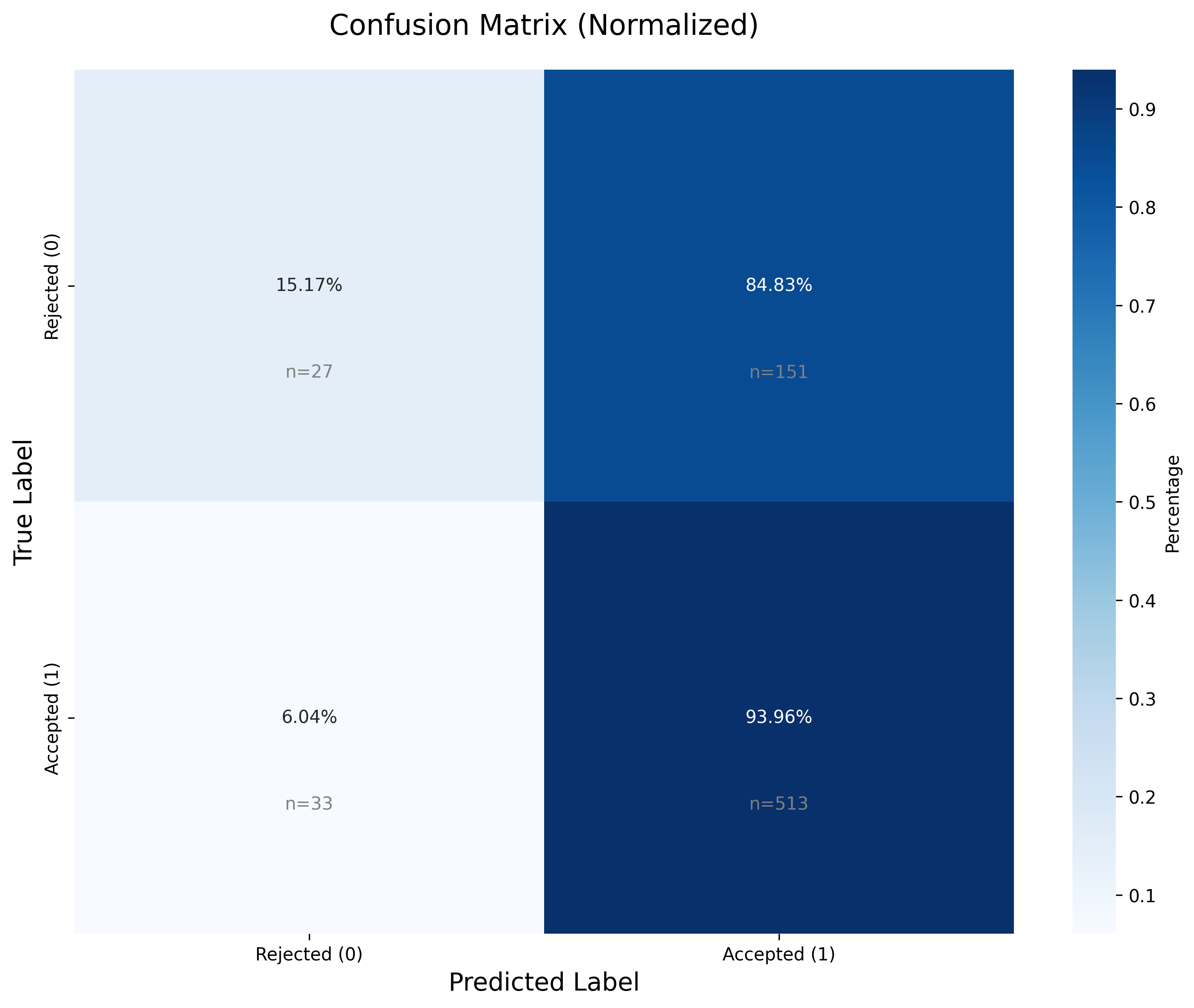}
    \caption{RF}
  \end{subfigure}\hfill
  \begin{subfigure}[t]{0.24\textwidth}
    \centering
    \includegraphics[width=\linewidth]{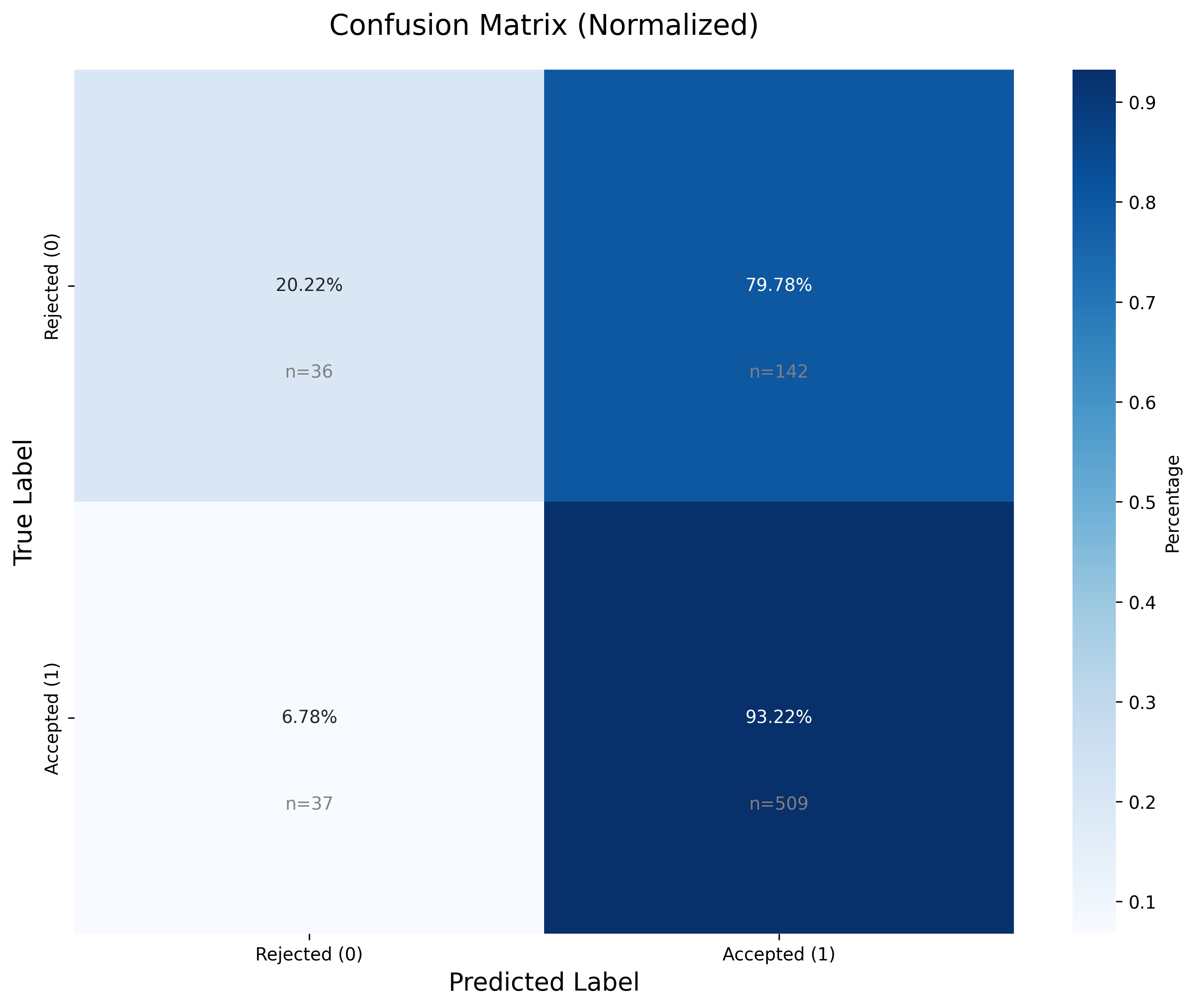}
    \caption{XGBoost}
  \end{subfigure}
  \caption{Row-normalized confusion matrices of four predictors.}
  \Description{The confusion matrices show the MLP abtains the lowest false alarm rate, making it best for reducing the WDT}
  \label{fig:predictor:cms}
\end{figure*}

This operating-point difference has direct implications for our \emph{stop-at-first-predicted-rejection} drafting policy.
Under this policy, \emph{false positives} (predicting Accepted when the token would be Rejected) are the primary cause of over-drafting and WDT,
because they allow the device to continue drafting past the first true rejection.
As shown in Figure.\ref{fig:predictor:cms}, the tree-based predictors incur very high false-positive rates on rejected tokens
(FPR $=0.80$--$0.86$), meaning that most truly rejected tokens are incorrectly labeled as accepted (Table~\ref{tab:predictor:cm_appendix}),
which would translate into frequent over-drafting whenever a rejection occurs early.
Conversely, the MLP reduces FPR to 0.425 (a $\sim$2$\times$ reduction versus the tree-based baselines), thus better aligning with the objective of minimizing wasted drafting.
Meanwhile, the tree-based models yield low false negatives on accepted tokens (FNR $=0.06$--$0.07$ implied by Rec$_1$),
which favors longer speculative windows and higher raw drafting utilization when rejections are rare.
Therefore, the choice of predictor is fundamentally a \emph{waste-vs-coverage} trade-off: minimizing WDT calls for low FPR/high specificity (favoring the MLP),
whereas maximizing speculative coverage prioritizes high Rec$_1$ (favoring the tree-based models). In our case, we pick MLP model for lower WDT, as we motivate in Section.\ref{sec:motivations_wdt}.

\paragraph{Computation overhead on edge devices.}
The predictor is invoked once per drafted token; thus its per-inference latency is critical.
We benchmark the four predictor candidates on two representative edge devices (Raspberry Pi~5 and Raspberry Pi~4B)
and report both single-sample latency (1,000 trials).
Table~\ref{tab:latency-single} reports single-sample statistics. The results show that XGBoost and the MLP achieve sub-millisecond decisions on RPi~5 and only a few milliseconds on RPi~4B,
while random forests incur tens to over one hundred milliseconds, making them unsuitable for per-token control loops despite strong accuracy.

\begin{table*}[t]
  \centering
  \small
  \caption{Single-sample inference latency (ms) for predictor models on edge devices.}
  \label{tab:latency-single}
  \begin{tabular}{@{}lrrrrrrr@{}}
    \toprule
    \multirow{2}{*}{\textbf{Device / Model}} & \multicolumn{7}{c}{\textbf{Latency statistics (ms)}} \\
    \cmidrule(l){2-8}
    & Mean & Median & Std & Min & Max & P95 & P99 \\
    \midrule
    \multicolumn{8}{l}{\textbf{Raspberry Pi 5}} \\
    XGBoost     & 0.3484 & 0.3405 & 0.0502 & 0.3302 & 1.3713 & 0.3595 & 0.6226 \\
    LightGBM    & 0.6827 & 0.6756 & 0.0647 & 0.6675 & 2.5907 & 0.6933 & 0.7866 \\
    RandomForest& 42.6230& 44.4504& 4.3045 & 32.5311& 46.4203& 45.4629& 45.7573 \\
    MLP         & 0.4590 & 0.4552 & 0.0556 & 0.4462 & 2.1973 & 0.4733 & 0.4831 \\
    \midrule
    \multicolumn{8}{l}{\textbf{Raspberry Pi 4B}} \\
    XGBoost     & 1.6221 & 1.3876 & 0.5883 & 1.3376 & 3.9257 & 3.1273 & 3.1645 \\
    LightGBM    & 2.6048 & 2.5467 & 0.2668 & 2.4819 & 6.9231 & 3.2125 & 3.3916 \\
    RandomForest& 114.4152&102.6887&24.6844& 96.8155&183.0372&169.2718&174.3341 \\
    MLP         & 2.0381 & 2.0352 & 0.1863 & 1.9750 & 7.8550 & 2.0735 & 2.0945 \\
    \bottomrule
  \end{tabular}
\end{table*}

\paragraph{Takeaway for deployment.}
Considering the overall accuracy, especially false alarm rate, and the inference overhead, we adopt MLP as the predictor in the paper.

\section{Profiling and Validation of the Verification Time Estimator}
\label{sec:appendix:profiler}

To obtain the coefficients in our latency model, we design a systematic profiling methodology that captures the diverse workload characteristics encountered in batch verification. The profiling process consists of three phases: experimental data collection, model fitting with statistical validation, and generalization testing on unseen configurations.

\paragraph{Experimental Design.}
We construct a comprehensive dataset of 173 configurations spanning five categories to ensure coverage of both compute-bound and memory-bound regimes. The training set comprises 123 configurations: (1) 25 compute-bound deterministic configurations with total tokens ranging from 1200 to 2000 and varying batch patterns (single, split-2, split-4, varied), all with $L_{\text{cached}} = 0$; (2) 48 memory-bound deterministic configurations with $L_{\text{new}} \in \{1, 5, 10, 20, 50, 100\}$, $L_{\text{total}} \in \{500, 1000, 1500, 2000\}$, and batch sizes of 1 and 4, featuring high caching ratios where $L_{\text{cached}} = L_{\text{total}} - L_{\text{new}}$; (3) 15 compute-bound random configurations with high $L_{\text{new}} \geq 1200$; (4) 15 memory-bound random configurations with small $L_{\text{new}}$ and large $L_{\text{total}}$; and (5) 20 mixed configurations containing both cached and uncached requests within the same batch. The test set contains 50 completely unseen configurations generated with different random seeds to ensure true generalization testing.

\paragraph{Data Collection.}
We deploy Qwen3-32B on an NVIDIA A100 GPU with prefix caching enabled via vLLM. For each configuration, we perform 10 warmup iterations (discarded) followed by 3 independent measurement runs. To prevent cache pollution across runs---a critical issue we identified where prompt reuse caused over 60\% artificial speedup---we generate unique prompts for each run using timestamp-based seeding:
\begin{verbatim}
unique_prefix = f"[RUN_{run_idx}_REQ_{req_idx}_{time.time()}]"
\end{verbatim}
We verify prompt uniqueness before each measurement and raise an exception if duplicates are detected. For each configuration, we record the mean, standard deviation, minimum, maximum, and median latency across the 3 runs, targeting a coefficient of variation below 5\% for measurement quality.

\paragraph{Model Fitting.}
We employ OLS linear regression via \texttt{\path{sklearn.linear_model.LinearRegression}} to fit the model parameters. The feature matrix $\mathbf{X} \in \mathbb{R}^{n \times 3}$ contains the columns $[N_{\text{linear}}, N_{\text{interactions}}, N_{\text{cached}}]$, and the target vector $\mathbf{y} \in \mathbb{R}^{n}$ contains the measured latencies. We fit exclusively on the 123 training samples and evaluate on the 50 held-out test samples. To quantify parameter uncertainty, we perform 1000 bootstrap iterations, resampling the training set with replacement and refitting the model each time. The 95\% confidence intervals are computed as the 2.5th and 97.5th percentiles of the bootstrap parameter distributions.

\paragraph{Validation Metrics.}
We evaluate model quality using multiple metrics: coefficient of determination ($R^2$), adjusted $R^2$, Root Mean Squared Error (RMSE), MAE, and MAPE. Additionally, we perform 5-fold cross-validation on the training set to assess model stability. The generalization quality is measured by the $R^2$ difference between training and test sets, with a threshold of $< 0.05$ indicating excellent generalization.

\paragraph{Results.}
Table~\ref{tab:latency_params} presents the fitted parameters with their 95\% confidence intervals. The linear operation coefficient $a = 3.314 \times 10^{-5}$ sec/token (33.14 $\mu$s/token) reflects the per-token cost for embeddings, projections, and FFN layers. The attention compute coefficient $b_{\text{compute}} = 3.450 \times 10^{-8}$ sec/interaction (0.0345 $\mu$s/interaction) captures the cost per query-key dot product. The memory read coefficient $b_{\text{read}} = 4.620 \times 10^{-6}$ sec/cached\_token (4.62 $\mu$s/token) represents the HBM bandwidth cost for loading cached KV states. The constant overhead $c = 14.86$ ms accounts for kernel launch and scheduling latency.

\begin{table}[t]
\centering
\caption{Fitted latency model parameters with 95\% confidence intervals}
\label{tab:latency_params}
\begin{tabular}{lrrl}
\toprule
Parameter & Value & 95\% CI & Unit \\
\midrule
$a$ & $3.314 \times 10^{-5}$ & $[3.12, 3.49] \times 10^{-5}$ & sec/token \\
$b_{\text{compute}}$ & $3.450 \times 10^{-8}$ & $[3.35, 3.67] \times 10^{-8}$ & sec/interaction \\
$b_{\text{read}}$ & $4.620 \times 10^{-6}$ & $[3.95, 5.27] \times 10^{-6}$ & sec/cached\_token \\
$c$ & $1.486 \times 10^{-2}$ & $[1.30, 1.69] \times 10^{-2}$ & sec \\
\bottomrule
\end{tabular}
\end{table}

\begin{figure*}[t!]
    \centering
    \includegraphics[scale=0.4]{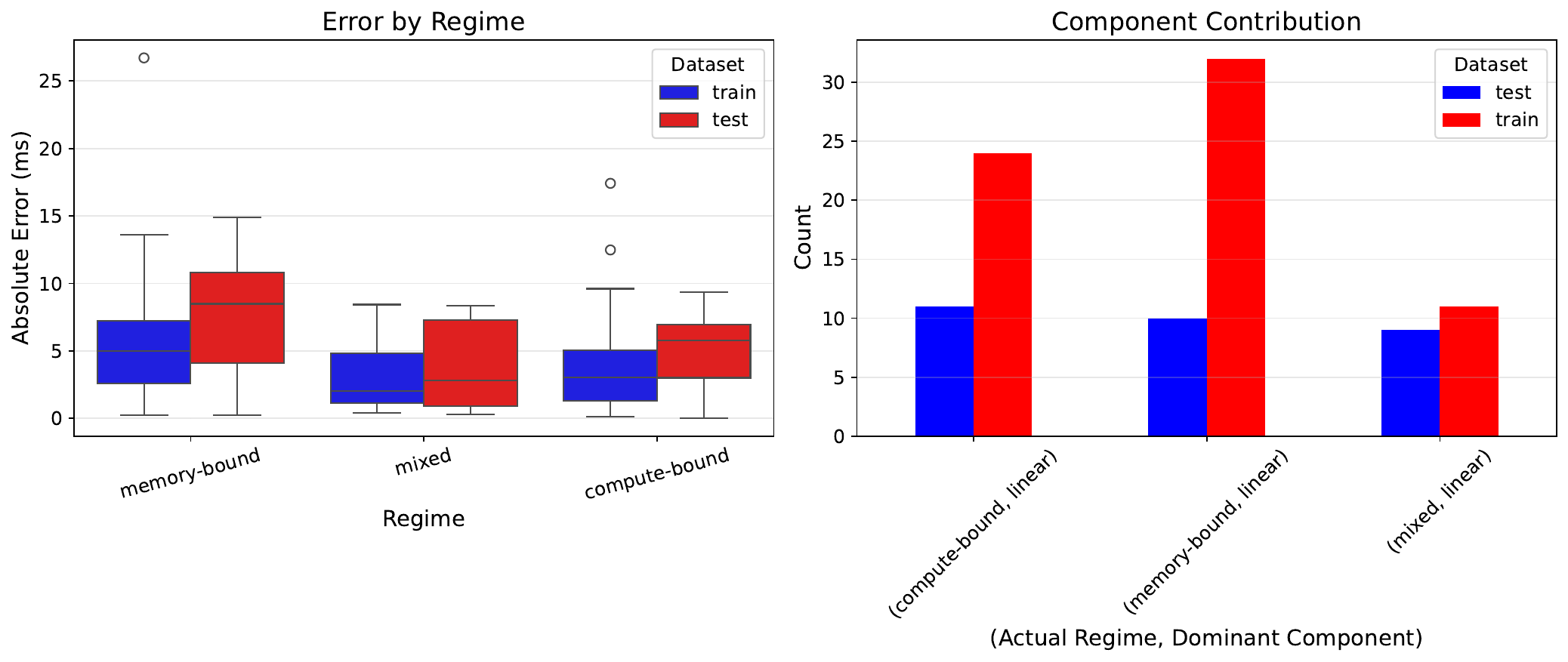}
    \caption{Error breakdown of the additive latency predictor across workload regimes}
    \Description{(Left) Absolute prediction error under memory-bound, mixed, and compute-bound regimes. (Right) Sample composition by (regime, dominant component), illustrating coverage of heterogeneous verification workloads.}
    \label{fig:additive_component_analysis}
\end{figure*}

\paragraph{Error stratification and workload coverage (Figure.~\ref{fig:additive_component_analysis}).}
Figure.~\ref{fig:additive_component_analysis} further decomposes estimator behavior by workload regime. The left panel shows absolute-error distributions stratified into compute-bound, memory-bound, and mixed settings; errors remain consistently small across regimes, with only a modest widening in the more memory-sensitive cases. This pattern is expected because memory-bound batches are more exposed to cache-read variability and kernel-level effects that are harder to model with a single linear coefficient. The right panel summarizes how the evaluated configurations distribute across (regime, dominant component), confirming that both training and test sets cover diverse operating regions rather than concentrating on a single regime. Together, these results indicate the estimator is not only accurate on average, but also robust across the heterogeneous verification workloads that arise in speculative serving.

\paragraph{Component Analysis.}
To understand the relative importance of each latency component, we analyze their contributions across different workload types. For compute-bound configurations (high $L_{\text{new}}$, low $L_{\text{cached}}$), the $N_{\text{interactions}}$ term dominates, contributing over 70\% of the predicted latency. For memory-bound configurations (low $L_{\text{new}}$, high $L_{\text{cached}}$), the $N_{\text{cached}}$ term becomes significant, contributing 30--50\% of the total latency. The linear term $N_{\text{linear}}$ provides a consistent baseline contribution proportional to batch complexity, while the constant overhead $c$ is most significant for small batches with short sequences. This decomposition validates our additive model structure and confirms that the three workload metrics capture orthogonal resource costs.

\textcolor{black}{\section{Output quality comparison between \acronym and traditional auto-regressive decoding}
\label{sec:appendix:quality}
Table~\ref{tab:model_quality} compares the output quality of \acronym and traditional autoregressive (AR) decoding across three datasets: GSM8K~\cite{cobbe2021gsm8k}, MMLU~\cite{hendrycks2021mmlu}, and ARC-Challenge~\cite{clark2018arc}. From the results in Table~\ref{tab:model_quality}, we observe that \acronym achieves nearly identical quality and accuracy to AR decoding on all three benchmarks. This is expected, since \acronym adopts a lossless speculative decoding verification scheme that preserves the target model's output distribution, ensuring that the final predictions remain consistent with those of standard AR decoding.}

\begin{table}[t]
\centering
{\color{black}
\caption{Output quality (accuracy \%) of target models across three benchmark datasets. Each model is evaluated three times with 200 samples per run (temperature = 0.6).}
\label{tab:model_quality}
\begin{tabular}{l c ccc c}
\toprule
\textbf{Model} & \textbf{System} & \textbf{GSM8K} & \textbf{MMLU} & \textbf{ARC-C} \\
\midrule
\multirow{2}{*}{Qwen3-14B}
  & \acronym    & 91.50 & 75.00 & 92.50\\
  & AR    & 92.50 & 77.00 & 94.00\\
\midrule
\multirow{2}{*}{Qwen3-32B}
  & \acronym    & 93.00 & 79.50 & 94.50 \\
  & AR    & 94.00 & 77.00 & 95.00 \\
\midrule
\multirow{2}{*}{Llama-3.1-70B}
  & \acronym    & 96.00 & 80.50 & 90.00 \\
  & AR    & 95.50 & 80.00 & 89.50 \\
\bottomrule
\end{tabular}
}
\end{table}

\end{document}